\documentclass[reprint,superscriptaddress,amsmath,amssymb,aps]{revtex4-2}
\usepackage{graphicx}
\usepackage{dcolumn}
\usepackage{bm}
\usepackage[colorlinks,linkcolor=blue,citecolor=blue,urlcolor = blue]{hyperref}
\usepackage{multirow} 
\usepackage{siunitx}
\begin{document}

\title{Mode-Resolved Light Scattering Recovers Polymer Thermodynamics \\ in Solutions with Trace Large-Mass Scatterers}

\author{Noriaki Mizumoto}
\thanks{These authors contributed equally: N. Mizumoto, T. Yasuda}
\affiliation{Division of Soft Matter, Graduate School of Life Science, Hokkaido University, Sapporo 060-0810, Japan.}
\author{Takashi Yasuda}
\thanks{These authors contributed equally: N. Mizumoto, T. Yasuda}
\affiliation{Faculty of Advanced Life Science, Hokkaido
University, Sapporo 001-0021, Japan.}
\author{Hiroshi Hinou}
\affiliation{Faculty of Advanced Life Science, Hokkaido
University, Sapporo 001-0021, Japan.}
\author{Xiang Li}
\email[Contact author:]{\,
x.li@sci.hokudai.ac.jp}
\affiliation{Faculty of Advanced Life Science, Hokkaido
University, Sapporo 001-0021, Japan.}

\date{\today}

\begin{abstract}
Light scattering provides direct access to polymer conformations and thermodynamics. 
However, trace large-mass scatterers such as aggregates and nanobubbles form unavoidably in polymer solutions and dominate the scattered intensity, obscuring the intrinsic polymer signal. We demonstrate that resolving the static scattering intensity by molecular mobility cleanly separates the polymer and large-scatterer contributions. 
Applying this mode-resolved analysis to aqueous poly(ethylene glycol) solutions, we isolate the polymer scattering even when these scatterers account for more than 90 \% of the total intensity. 
The resolved polymer intensity recovers the universal osmotic equation of state from the dilute to the semidilute regime over 288 to 358 K. 
This approach establishes a reliable basis for measuring the thermodynamics of interacting macromolecules in solutions where irreproducible large-mass scatterers have precluded quantitative analysis.

\end{abstract}
\maketitle
Scattering probes internal structure through the interference of scattered waves, resolving features well below the diffraction limit.
Light scattering carries a further capability for soft matter: in static light scattering (SLS), the absolute intensity extrapolated to zero angle measures the osmotic compressibility, giving thermodynamic access to interactions at the molecular scale \cite{Hulst1957_LS,BerneandPecora1976_LS,bohren2008absorption}.
In polymer solutions, SLS yields key thermodynamic and structural properties \cite{Berry1966_Berryplot,Kawaguchi1996_LS,Devanand1991_LS,Okumono1998_LS,Miyaki1978_LS,Outher1950_LS,Yamamoto1971_LS,Hasse1995_LS,Fukuda1974_LS,Noda1981_EOS,Adam1991_LS,Merkle1993_LS,Roovers1995_LS,Schaefer1983_LS,venohr1998static,Strazielle1968_thetaTempSLS,William1983_SLSPEOTemp20to90,Kinugasa1994_slowmode}.
However, solutions almost inevitably contain large-mass scatterers such as polymer aggregates, nanobubbles, and dust, reported for biological macromolecules \cite{Pecora1988_natural,Zhang2010_natural,Li2011_lifescience}, polyelectrolytes \cite{Sedlak1996_ele,Buhler2000_ele,Forster1990_ele}, and neutral synthetic polymers \cite{Brown1985_slowmode,Alami1996_slowmode,Wang2015_slowmode,Polverari1996_slowmode,Ho2002_slowmode,Sun1996_slowmode,Durval2000_slowmode,Raspaud1994_theory,Stepanek1998_theory,Kanao2003_slowmode,Yuan2006_slowmode,Li2010_slowmode,Wang2016_slowmode,Bica1997_slowmode,Brown1987_slowmode,Fang1990_slowmode_P,Fang1990_slowmode_M,Brown1986_slowmode,Brown1984_slowmode,Zhou1989_slowmode}.
The scattering intensity scales linearly with the number density $N$ but quadratically with the mass $M$, $I \propto NM^2$ \cite{bohren2008absorption}, so even a trace mass fraction of them scatters intensely.
Because SLS records only the sum over all scatterers, this contribution cannot be separated from that of the polymers.

\begin{figure}[b!]
\centering
\includegraphics[width=\linewidth]{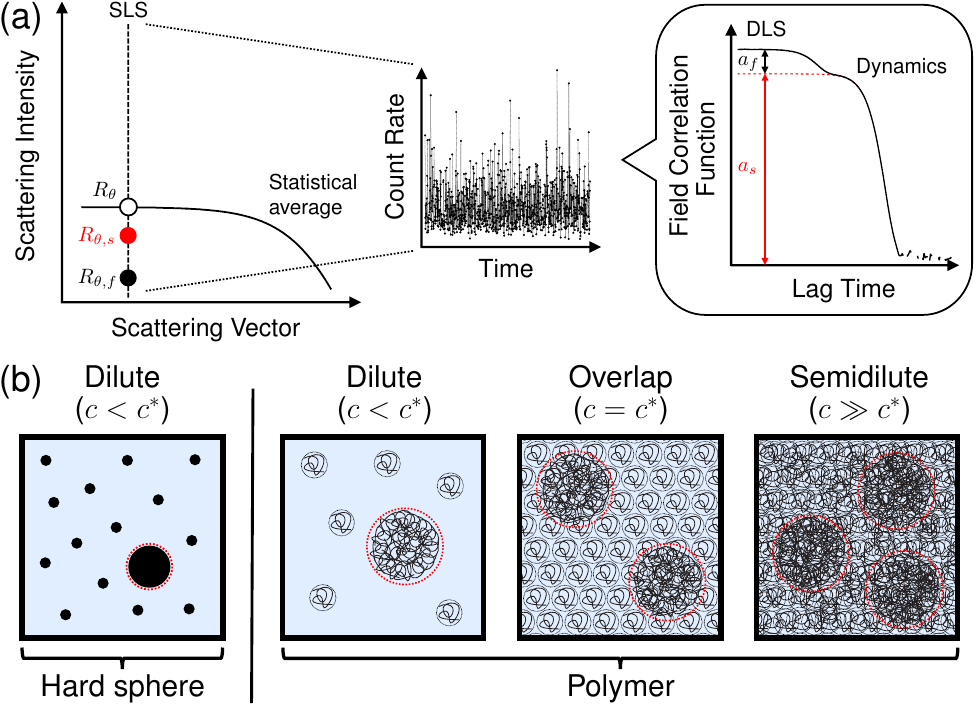}
\caption{(a) Overview of the mode-resolved analytical framework in light scattering, illustrated for polymer solutions containing large-mass scatterers.
The scattering intensity of all solutes $R_\theta$, a sum over modes, is resolved into the intrinsic polymer (fast) and slow components ($R_{\theta,f}$ and $R_{\theta,s}$), using the amplitudes ($a_f$ and $a_s$) of the corresponding modes in the field correlation function.
(b) Schematic of the fast and slow modes in hard-sphere and polymer solutions.
The small black and large red scatterers represent the fast and slow modes, respectively.
Hard-sphere systems are confined to the dilute regime, whereas polymer chains can overlap beyond the overlap concentration $c^*_{A_2}$ and fill space in the semidilute regime.
}
\label{fig:Schematic}
\end{figure}

The straightforward remedy is removal, through careful sample preparation, repeated filtration, and degassing \cite{Wang2015_slowmode,Kinugasa1994_slowmode,Ho2002_slowmode}.
Such protocols demand considerable effort and fail for viscous solutions or polymers with labile bonds.
An alternative is to separate the polymer and large-scatterer contributions by their difference in mobility, using dynamic light scattering (DLS).
Whereas SLS averages the intensity over time, DLS tracks its fluctuations and sorts the scattered intensity by relaxation rate \cite{Pecora1988_natural,Zhang2010_natural,Li2011_lifescience, Sedlak1996_ele,Buhler2000_ele,Forster1990_ele,Brown1985_slowmode,Alami1996_slowmode,Wang2015_slowmode,Polverari1996_slowmode,Ho2002_slowmode,Sun1996_slowmode,Durval2000_slowmode,Raspaud1994_theory,Stepanek1998_theory,Kanao2003_slowmode,Yuan2006_slowmode,Li2010_slowmode,Wang2016_slowmode,Bica1997_slowmode,Brown1987_slowmode,Fang1990_slowmode_P,Fang1990_slowmode_M,Brown1986_slowmode,Brown1984_slowmode,Zhou1989_slowmode}.
The total intensity can thus be resolved into the contribution of each relaxation mode [Fig.~\ref{fig:Schematic}(a)] \cite{Raspaud1994_theory}.

This mode-resolved framework has been applied to hard-sphere suspensions, which are necessarily confined to the dilute regime \cite{Pusey1982_hardsphere,Ju1992_hardsphere}.
For polymer solutions, evidence is limited to dilute mixtures of latex and polymer \cite{Shibayama2006_hardsphere} and to one dilute and one semidilute concentration in a single aggregate-containing system \cite{Kanao2003_slowmode}.
Two questions therefore remain open: whether the framework holds continuously across the dilute-to-semidilute crossover, where polymer coils overlap and become spatially indistinguishable [Fig.~\ref{fig:Schematic}(b)], and whether it holds as the solvent quality, and hence the second virial coefficient $A_2$, is varied.

In this Letter, we test the framework on aqueous poly(ethylene glycol) (PEG) solutions, a system notorious for anomalous excess scattering at small angles \cite{Hammouda2004_slowmode} and a slow relaxation mode in DLS \cite{Brown1985_slowmode,Alami1996_slowmode,Polverari1996_slowmode,Wang2015_slowmode,Ho2002_slowmode,Brown1984_slowmode}.
Combining SLS and DLS, we resolve the scattering intensity into distinct relaxation modes and extract the thermodynamics of the solvated polymers from good-solvent to near-theta conditions by varying the temperature.
With the universal osmotic equation of state \cite{Noda1981_EOS,Higo1983_EOS,Cloizeaux1975_EOS,Cloizeaux1982_EOS,Ohta1982_EOS,Ohta1983_EOS} as a stringent reference, we verify that the framework applies continuously from the dilute to the semidilute regime and across solvent qualities.

\textit{Theoretical Framework}---For a polymer solution containing large-mass scatterers, the field correlation function $g^{(1)}(\tau)$ measured by DLS typically exhibits two distinct relaxation modes:
\begin{equation}
g^{(1)}(\tau) = a_{f} \exp(-\Gamma_{f}\tau) + a_{s} \exp(-\Gamma_{s}\tau),
\label{eq:g1}
\end{equation}
where $\Gamma_{f}$ and $\Gamma_{s}$ are the relaxation rates of the fast and slow modes, respectively; $a_{f}$ and $a_{s}$ are their fractional intensity contributions ($a_f + a_s = 1$); and $\tau$ is the lag time.
For Brownian motion, each relaxation rate gives a translational diffusion coefficient $D = \Gamma/q^2$ \cite{Hulst1957_LS,BerneandPecora1976_LS}.
Here $q \equiv (4\pi n/\lambda_0)\sin(\theta/2)$ is the scattering vector magnitude, with $n$ the refractive index of the solution, $\lambda_0$ the incident wavelength in vacuum, and $\theta$ the scattering angle.
Solvated polymer chains diffuse far faster than any large-mass scatterer, so the fast mode carries the polymer dynamics and the slow mode carries the large-mass scatterers.

The Rayleigh ratio (absolute scattering intensity) of the solvated polymers then follows from the overall solute scattering \cite{Raspaud1994_theory},
\begin{equation}
R_{\theta,f} = a_{f}(\theta)\,R_{\theta},
\label{eq:Rayleigh_separation}
\end{equation}
and likewise for the slow mode, $R_{\theta,s} = a_{s}(\theta)R_{\theta}$ [see the Supplemental Material (SM), Sec.~S1 \cite{supplement}, for the derivation of Eqs.~(\ref{eq:g1}) and (\ref{eq:Rayleigh_separation})].
Here, $R_\theta$ represents the Rayleigh ratio of all solutes, including the large-mass scatterers, defined as the absolute scattering of the polymer solution minus the solvent contribution.
Equation~(\ref{eq:Rayleigh_separation}) requires only that the modes be separated in relaxation rate and that their concentration fluctuations be uncorrelated. Under these conditions it generalizes to any number of components, $R_{\theta,i} = a_i(\theta)R_\theta$.

Fluctuation theory relates the resolved polymer intensity at zero angle to the osmotic compressibility $\partial c/\partial\Pi$ \cite{BerneandPecora1976_LS}, conventionally written in reciprocal form:
\begin{equation}
\frac{Kc}{R_{\theta=0,f}}=\frac{1}{RT}\left(\frac{\partial\Pi}{\partial c}\right),
\label{eq:FT}
\end{equation}
where $K$ is the optical constant (see SM, Sec.~S2 \cite{supplement} for its determination), $c$ is the polymer mass concentration, $R$ is the gas constant, and $T$ is the absolute temperature.

For neutral and flexible linear polymer solutions in good solvents, the osmotic pressure obeys a universal equation of state (EOS) across the dilute and semidilute regimes, hereafter the universal EOS \cite{Noda1981_EOS,Higo1983_EOS,Cloizeaux1975_EOS,Cloizeaux1982_EOS,Ohta1982_EOS,Ohta1983_EOS} (see End Matter for details).
In the dilute regime ($c < c^*_{A_2}$), the classical virial expansion holds for $\partial \Pi / \partial c$ \cite{Flory1953_polymer}:
\begin{equation}
\frac{1}{RT}\left(\frac{\partial\Pi}{\partial c}\right)=\frac{1}{M_w}+2A_2c+3A_3c^2+\cdots,
\label{eq:VE}
\end{equation}
where $M_w$ is the weight-average molar mass to which light scattering is sensitive and $A_3$ is the third virial coefficient.
The coefficient $A_2$ measures pairwise interactions between chains: positive in a good solvent, zero at the theta condition.
The polymer overlap concentration, which marks the crossover between the dilute and semidilute regimes, is defined thermodynamically as $c^{*}_{A_2} \equiv 1/(A_2M_w)$.
In the semidilute regime ($c \gg c^{*}_{A_2}$), $\partial \Pi / \partial c$ approaches a scaling law in a good solvent \cite{Cloizeaux1975_EOS,Gennes1979_polymer}:
\begin{equation}
\frac{\partial \Pi}{\partial c} \sim c^{1/(3\nu-1)}.
\label{eq:scaling}
\end{equation}
The excluded-volume exponent $\nu \approx 0.588$ gives $1/(3\nu-1) \approx 1.31$ \cite{Flory1953_polymer,Gennes1979_polymer}.

\textit{Materials and Methods}---Linear methoxy-PEG-OH [$\mathrm{CH_3O}$--$(\mathrm{CH_2}$--$\mathrm{CH_2}$--$\mathrm{O})_n$--$\mathrm{H}$] with nominal number-average molar mass $M_{n,\mathrm{nom}} \approx 40$ kg$/$mol and polydispersity $\mathrm{PDI}_{\mathrm{nom}} \approx 1.2$ (XIAMEN SINOPEG BIOTECH Co., Ltd., China) was dissolved at $c = 0.4$--$100$ g$/$L in pure water at room temperature.
Matrix-assisted laser desorption/ionization time-of-flight mass spectrometry (MALDI-TOF MS) gave $M_{n,\mathrm{MALDI}} \approx 43.2$ kg$/$mol, $M_{w,\mathrm{MALDI}} \approx 43.5$ kg$/$mol, and $\mathrm{PDI}_{\mathrm{MALDI}} \approx 1.01$ (see SM, Sec.~S3 \cite{supplement} for details).
Throughout, $c$ denotes the polymer mass per solvent volume, a definition that extends the universal equation of state to higher concentrations \cite{yasuda2020_universal}.
The solutions were filtered through $0.22$ $\mathrm{\mu}$m hydrophilic syringe filters (Membrane Solutions, USA).

The time-averaged scattering intensity and the intensity correlation function $g^{(2)}(\tau)$ of the PEG solutions and pure water were measured using a custom-built apparatus equipped with a vertically polarized solid-state laser ($\lambda_0 = 660$ nm, $500$ mW, Cobolt, Sweden).
Scattered photon counts $I(t)$ were recorded for $60$ s at $\theta = 50^{\circ}$--$150^{\circ}$ in $10^{\circ}$ steps and over $288$--$358$ K in $10$ K steps.
The total Rayleigh ratio of the solution was obtained with toluene as the reference, $R_{\theta, \mathrm{solution}} = (n_{\mathrm{solution}}/n_{\mathrm{tol}})^2 \left( \langle I(t) \rangle / \langle I_\mathrm{tol}(t) \rangle \right) R_{\theta,\mathrm{tol}}$, where $\langle\cdots\rangle$ denotes the time average.
Here $n_{\mathrm{solution}}$ and $n_{\mathrm{tol}}$ are the refractive indices of the solution and toluene, and $I_{\mathrm{tol}}$ and $R_{\theta,\mathrm{tol}}$ are the scattering intensity and Rayleigh ratio of toluene at $298$ K \cite{Alexey2021_tolueneRayleigh}.
The Rayleigh ratio of all solutes $R_\theta$ was then evaluated by subtracting the volume-fraction-corrected solvent baseline: $R_\theta = R_{\theta,\mathrm{solution}} - (1-\phi)R_{\theta,\mathrm{water}}$ (see SM, Sec.~S1 \cite{supplement} for details), where $R_{\theta,\mathrm{water}}$ is the Rayleigh ratio of pure water (see SM, Sec.~S4 \cite{supplement} for the measured data).
Here, $\phi$ is the polymer volume fraction, calculated from $c$ using the mass densities of PEG ($\approx 1125$ kg$/$m$^3$) and pure water ($\approx 1000$ kg$/$m$^3$).
The intensity correlation function was computed as $g^{(2)}(\tau) \equiv \langle I(0)I(\tau) \rangle / \langle I(0) \rangle^2$, using a digital correlator (LS Instruments, Switzerland).
The field correlation function follows from the Siegert relation, $g^{(2)}(\tau) = 1 + \beta_{\mathrm{eff}}|g^{(1)}(\tau)|^2$, where the effective coherence factor $\beta_{\mathrm{eff}}$ absorbs the instrumental coherence, detector noise, dead time, and solvent scattering (see SM, Sec.~S1 \cite{supplement} for details).

\begin{figure}[t!]
\centering
\includegraphics[width=\linewidth]{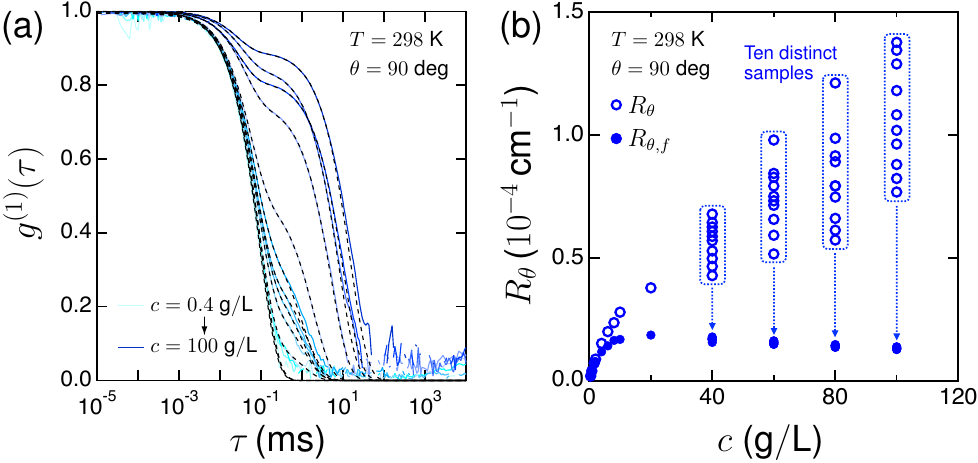}
\caption{Mode-resolved analysis to extract $R_{\theta,f}$ from the measured $R_{\theta}$, at $T = 298$ K and $\theta = 90^{\circ}$ in aqueous PEG solutions.
(a) The $g^{(1)}(\tau)$ profile for $c = 0.4$--$100$ g$/$L (different colors).
Black dashed curves are least-squares fits of Eq.~(\ref{eq:g1}) at each $c$.
(b) The $c$ dependence of $R_{\theta,f}$ (filled circles) resolved from $R_{\theta}$ (open circles) via Eq.~(\ref{eq:Rayleigh_separation}) with $a_f$ from (a).
For $c=40$--$100$ g$/$L, ten distinct samples were prepared under identical conditions (blue dashed enclosure).}
\label{fig:separation}
\end{figure}

\textit{Isolation of Solvated Polymer Scattering}---We resolved $R_{\theta,f}$ from the overall solute scattering $R_\theta$ by fitting the measured $g^{(1)}(\tau)$ with Eq.~(\ref{eq:g1}). Figure~\ref{fig:separation}(a) shows representative $g^{(1)}(\tau)$ profiles for aqueous PEG solutions ($c=0.4$--$100$ g$/$L) at a fixed scattering angle and temperature ($T=298$ K and $\theta = 90^{\circ}$, see SM, Sec.~S5 \cite{supplement} for other $T$ and $\theta$).
As the polymer concentration increases, the slow mode becomes prominent and its relaxation time lengthens.
The origin of this slow mode, attributed variously to polymer aggregates and to nanobubbles, remains under debate \cite{Kinugasa1994_slowmode,Sun1996_slowmode,Polverari1996_slowmode,Durval2000_slowmode,Devanand1991_LS,Ho2002_slowmode,Hammouda2004_slowmode,Hammouda2002_slowmode,Brown1984_slowmode}.
Our analysis does not depend on which assignment is correct: it requires only that the scatterers be large enough to diffuse slowly.
The $g^{(1)}(\tau)$ profiles are accurately described by Eq.~(\ref{eq:g1}) [black dashed curves in Fig.~\ref{fig:separation}(a)], yielding the amplitudes $a_f$ and $a_s$, and their corresponding relaxation rates (see SM, Secs.~S6, S7, and S8 \cite{supplement} for summarized data).
Using Eq.~(\ref{eq:Rayleigh_separation}), $R_{\theta,f}$ is then resolved from the total solute intensity $R_\theta$ measured by SLS [Fig.~\ref{fig:separation}(b)] (see SM, Sec.~S9 \cite{supplement} for other $T$).

To rigorously test this approach, we prepared ten distinct samples per concentration in the regime where the slow mode is pronounced ($c = 40$--$100$~g$/$L), under identical conditions using the same batch of PEG powder.
During measurements, the scattering intensity of each sample remained temporally stable (see SM, Sec.~S10 \cite{supplement} for details).
However, the $R_{\theta}$ exhibits large variation, which likely stems from the uncontrollable formation of aggregates or nanobubbles in standard preparation protocols \cite{Ho2002_slowmode} [see SM, Sec.~S11 \cite{supplement} for $g^{(1)}(\tau)$ profiles].
Despite this erratic variation in $R_\theta$, the resolved polymer intensities $R_{\theta,f}$ (filled circles) collapse onto a single, consistent value at each concentration.
This convergence demonstrates the robustness of the mode-resolved analysis.

\begin{figure}[t!]
\centering
\includegraphics[width=\linewidth]{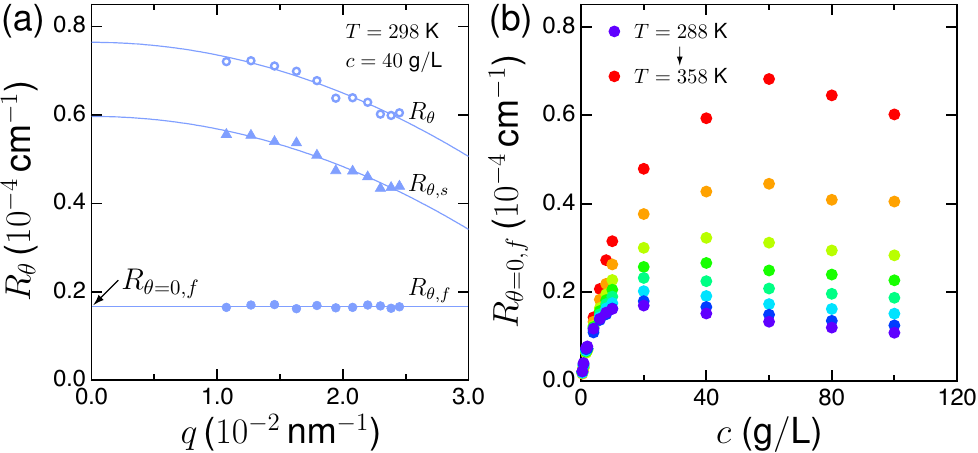}
\caption{Evaluation of $R_{\theta=0,f}$ by extrapolating $R_{\theta,f}$ to $\theta\to 0$.
(a) Angular profile of $R_{\theta}$ (open circles) for $c = 40$ g$/$L at $T = 298$ K over $\theta=50^{\circ{}}$--$150^{\circ{}}$ measured on a single sample, with constituents $R_{\theta,f}$ (filled circles) and $R_{\theta,s}$ (filled triangles).
The solid line is a constant fit to $R_{\theta,f}$.
Curves are Guinier fits to $R_{\theta}$ and $R_{\theta,s}$ (see SM, Sec.~S12 \cite{supplement} for details).
(b) The $c$ dependence of the resolved $R_{\theta=0,f}$ over $T = 288$--$358$ K.}
\label{fig:R0}
\end{figure}

\begin{figure*}[t!]
\centering
\includegraphics[width=\linewidth]{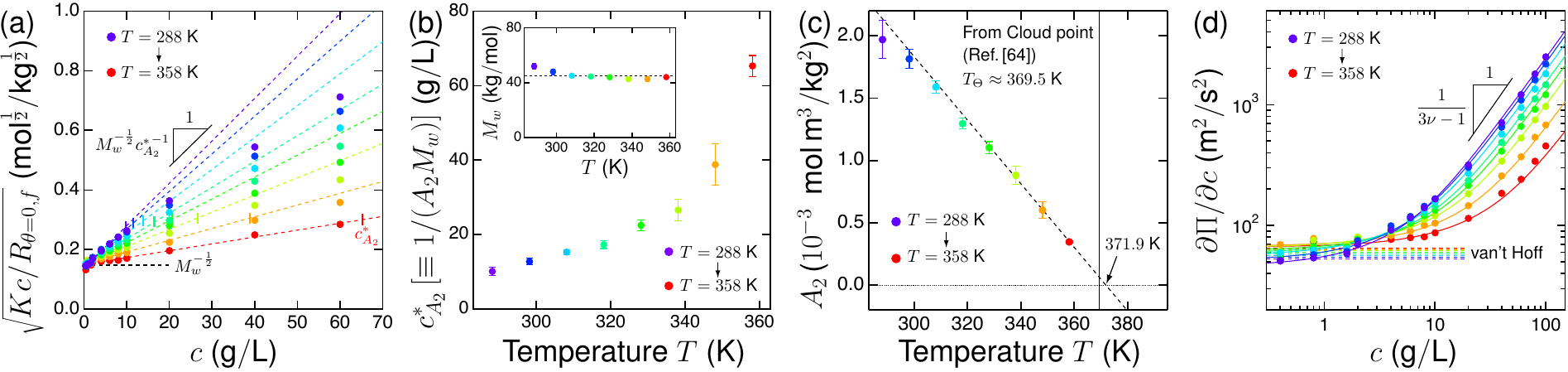}
\caption{Analysis of the resolved $R_{\theta=0,f}$ for $c^*_\mathrm{A_2}$, $M_w$, $A_2$, and $\partial\Pi/\partial c$ of aqueous PEG solutions over $T=288$--$358$ K.
(a) Berry plot using $R_{\theta,f}$.
(b) The $T$ dependence of $c^{*}_{A_2}$ from the Berry plot. Inset: That of $M_w$.
The black dashed line is the constant fit, giving $M_w= 45.5(11)$ kg$/$mol.
(c) The $T$ dependence of $A_2\,[\equiv1/(M_wc^*_{A_2})]$.
The black dashed line is the linear fit to the data in the vicinity of $A_2 = 0$, giving $T_\Theta = 371.9(9)$ K by extrapolation to $A_2=0$. 
The black solid line marks $T_\Theta = 369.5(11)$ K from the cloud point in Ref.~\cite{Fischer2000_thetaTempCPC}.
(d) The $c$ dependence of $\partial \Pi/\partial c$ from $R_{\theta=0,f}$ via Eq.~(\ref{eq:FT}).
Solid curves are predictions of the universal EOS for polymer solutions in good solvents [Eq.~(\ref{eq:UEOS}) in the End Matter]\cite{Noda1981_EOS,Higo1983_EOS,Cloizeaux1975_EOS,Cloizeaux1982_EOS,Ohta1982_EOS,Ohta1983_EOS} with the corresponding $M_w$ and $A_2$, asymptotic to the van't Hoff law for $c \to 0$ (dashed lines) and to the semidilute scaling law [Eq.~(\ref{eq:scaling})] with exponent $1/(3\nu-1)$ for $c\gg c^*_{A_2}$.}
\label{fig:Berry_dPdc}
\end{figure*}

To evaluate the zero-angle intensity, we examined the angular profiles of the resolved intensities.
Figure~\ref{fig:R0}(a) displays representative data at $c = 40$~g$/$L and $T = 298$~K (see SM, Sec.~S12 \cite{supplement} for other $c$ and $T$).
The slow mode $R_{\theta,s}$ shows a clear $q$ dependence, reflecting structures of several hundred nanometers, whereas the polymer mode $R_{\theta,f}$ stays approximately constant over the accessible $q$ range.
The plateau places both the coil size and the correlation length of the concentration fluctuations well below $q^{-1}$.
The zero-angle Rayleigh ratio of the polymer mode $R_{\theta=0,f}$ therefore follows from a zero-slope extrapolation to $\theta = 0$.

We applied the extrapolation to all measured data, providing a comprehensive map of $R_{\theta=0,f}$ across concentrations and temperatures [Fig.~\ref{fig:R0}(b)].
At any given temperature, as the concentration increases, $R_{\theta=0,f}$ initially increases, reaches a maximum, and then decreases,
reflecting the universal transition from the dilute to the semidilute regime.
The $R_{\theta=0,f}$ increases systematically with temperature. By Eq.~(\ref{eq:FT}), the increase comes from the explicit factor $T$ and from the osmotic compressibility, with a minor contribution from the optical constant.
The maximum in $R_{\theta=0,f}$, which marks the onset of suppressed concentration fluctuations, shifts to higher concentrations with increasing temperature.
Its position is fixed in units of $c/c^{*}_{A_2}$, so with $M_w$ independent of temperature the shift tracks $A_2$ alone [Eqs.~(\ref{eq:FT}) and (\ref{eq:VE})].

\textit{Derivation of Thermodynamic Parameters}---Having systematically resolved the intrinsic polymer scattering, we can now apply the classical thermodynamic framework to extract the molecular parameters, free from artifacts of the slow mode.
Substituting the virial expansion [Eq.~(\ref{eq:VE})] into Eq.~(\ref{eq:FT}) yields the zero-angle Berry relation: 
$\sqrt{Kc/R_{\theta=0,f}} \approx M_w^{-1/2}(1+c/c^{*}_{A_2})$,
which holds at low concentrations and is used to extract $M_w$ and $c^*_{A_2}$ from the scattering intensity \cite{Berry1966_Berryplot} (see SM, Sec.~S13 \cite{supplement} for the accuracy of the Berry plot). 
Figure~\ref{fig:Berry_dPdc}(a) shows Berry plots of the resolved $R_{\theta=0,f}$ at various temperatures.
Consistent with the use of a single polymer batch, the extracted weight-average molar mass $M_w$ remains constant across all temperatures, yielding an average value of $45.5(11)$~kg$/$mol [inset in Fig.~\ref{fig:Berry_dPdc}(b)].
Here, the numbers in parentheses indicate the standard error in the last digits.
This value agrees with $M_{w,\mathrm{MALDI}}\approx 43.5$ kg$/$mol within experimental uncertainty.
In contrast, the obtained $c^{*}_{A_2}$ increases with temperature, indicating a reduced excluded-volume effect on the polymer chains and a higher concentration required for two-body interactions to become significant [main panel in Fig.~\ref{fig:Berry_dPdc}(b)].

Combining these gives $A_2 = 1/(M_wc^{*}_{A_2})$, which decreases with temperature [Fig.~\ref{fig:Berry_dPdc}(c)].
Water therefore becomes a poorer solvent for PEG as the temperature rises, the signature of the lower critical solution temperature (LCST) behavior of this system \cite{Saeki1976_thetaTempLCST}.
Extrapolation to $A_2 = 0$ yields the $\Theta$ temperature $T_\Theta = 371.9(9)$~K, where the net two-body interaction between polymer chains vanishes.
Within experimental uncertainty, this value agrees with $T_{\Theta}=369.5(11)$~K [black solid line in Fig.~\ref{fig:Berry_dPdc}(c)] determined from the cloud point in Ref.~\cite{Fischer2000_thetaTempCPC} (see SM, Sec.~S14 \cite{supplement} for details).
The unresolved $R_\theta$, which retains the slow mode, instead gives $T_\Theta = 357.6(36)$~K (see SM, Sec.~S15 \cite{supplement} for details), well below the resolved value.
The gap of $14$~K shows that the slow mode must be identified and separated before any thermodynamic quantity is extracted.
Interestingly, previous light and X-ray scattering studies report $T_\Theta$ between $350$ and $376$~K \cite{William1983_SLSPEOTemp20to90,venohr1998static,Strazielle1968_thetaTempSLS,pedersen2005temperature} (see SM, Sec.~S14 \cite{supplement} for a summary), a spread that may well stem from unrecognized aggregates or nanobubbles.

\begin{figure}[b!]
\centering
\includegraphics[width=\linewidth]{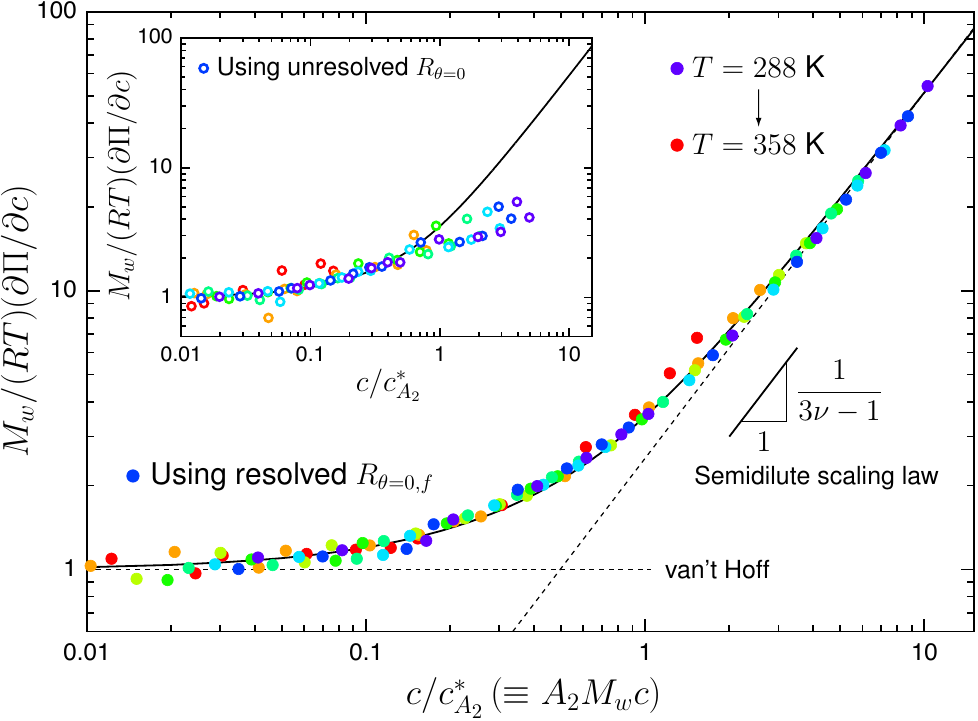}
\caption{Comparison of the resolved $R_{\theta=0,f}$ (filled circles, main panel) and the unresolved $R_{\theta=0}$ (open circles, inset) over $T = 288$--$358$ K with the universal EOS \cite{Noda1981_EOS,Higo1983_EOS,Cloizeaux1975_EOS,Cloizeaux1982_EOS,Ohta1982_EOS,Ohta1983_EOS} (black solid curves), asymptotic to the van't Hoff law for $c/c^*_{A_2} \to 0$ and to the semidilute scaling law [Eq.~(\ref{eq:scaling})] with exponent $1/(3\nu-1)$ for $c/c^*_{A_2} \gg 1$ (black dashed lines).}
\label{fig:PUEOS}
\end{figure}

Figure~\ref{fig:Berry_dPdc}(d) shows $\partial\Pi/\partial c$, derived from the resolved $R_{\theta=0,f}$ using Eq.~(\ref{eq:FT}), across a wide range of concentrations and temperatures. 
We test these data against the universal EOS \cite{Noda1981_EOS,Higo1983_EOS,Cloizeaux1975_EOS,Cloizeaux1982_EOS,Ohta1982_EOS,Ohta1983_EOS}, evaluated from Eq.~(\ref{eq:UEOS}) with the $M_w$ and $c^{*}_{A_2}$ obtained at each temperature (see End Matter).
Across the dilute to semidilute regime, $\partial\Pi/\partial c$ agrees quantitatively with the EOS at every temperature (solid colored curves).
In the dilute limit ($c\to 0$), $\partial\Pi/\partial c$ is asymptotic to the constant value dictated by the van't Hoff law ($\partial \Pi/\partial c = RT/M_w$), as indicated by the dashed lines for each temperature. 
The systematic upward shift of these lines with increasing temperature corresponds to the $RT$ term.
In the semidilute regime ($c \gg c^*_{A_2}$), 
$\partial\Pi/\partial c$ increases with concentration and approaches the good-solvent scaling of Eq.~(\ref{eq:scaling}) with the predicted exponent $1/(3 \nu - 1) \approx 1.31$. 
A slight deviation is observed at $T = 358$~K for $c > 80$~g$/$L, likely reflecting the reduced solvent quality indicated by $A_2$.

Normalizing $\partial\Pi/\partial c$ by $M_w/RT$ and the concentration $c$ by $c^{*}_{A_2}$ ($c/c^{*}_{A_2} \equiv A_2M_wc$) collapses all data onto the universal EOS \cite{Noda1981_EOS,Higo1983_EOS,Cloizeaux1975_EOS,Cloizeaux1982_EOS,Ohta1982_EOS,Ohta1983_EOS} over the full range of concentration and temperature (Fig.~\ref{fig:PUEOS}, main panel).
In contrast, the same plot built from the unresolved $R_\theta$ does not collapse, and the departure grows at $c/c^{*}_{A_2} > 1$, where the slow mode is strongest (Fig.~\ref{fig:PUEOS}, inset).
The universal relation is recovered from the resolved intensity and lost from the unresolved one. Extracting the thermodynamics of the solvated polymers therefore requires the mode-resolved analysis whenever a slow mode is present.

\textit{Concluding Remarks}---Through a combination of SLS and DLS on aqueous PEG solutions with temperature-dependent slow-mode fractions, our mode-resolved light scattering analysis [Eqs.~(\ref{eq:g1}) and (\ref{eq:Rayleigh_separation})] resolves the scattering intensity into distinct dynamic components.
The intrinsic scattering of the solvated polymers is recovered consistently, even when irreproducible slow-mode scatterers carry more than $90$\% of the total intensity.
The resolved intensity yields $\partial\Pi/\partial c$ in agreement with the universal EOS, which validates the method in the regime where the slow mode dominates.

Beyond this specific PEG system, our work establishes a general analytical framework for complex solutions. 
Conventional statistical averaging remains valid only when large-mass scatterers are absent ($a_s = 0$); otherwise, it severely biases the intrinsic thermodynamic analysis. Therefore, our findings underscore the necessity of verifying $a_f$ and $a_s$ prior to analysis and applying the mode-resolved approach whenever $a_s$ is nonzero. 
Because this framework is independent of the physical origin of the fluctuations, it can be broadly extended to diverse systems such as biological macromolecules \cite{Pecora1988_natural,Zhang2010_natural,Li2011_lifescience}, polyelectrolytes \cite{Sedlak1996_ele,Buhler2000_ele,Forster1990_ele}, coacervates formed via liquid-liquid phase separation \cite{Spruijt2013_coacervate}, and crystallizing polymers \cite{FerreDAmare1994_crystallized}.

\textit{Acknowledgments}---This work was supported by the 
JST ERATO Grant No.~JPMJER2401 to X.L., 
JST FOREST JPMJFR201Z to X.L., 
MEXT Project JPMXP1122714694 to X.L.,
J-PEAKS JPJS00420230001, Co-Creation Core for Soft Materials Aspiring Research \& Translation (C3-SMART) to X.L.
This work was also supported by the
Japan Society for the Promotion of Science (JSPS) through the Grants-in-Aid 
for Scientific Research (B) No.~JP22H02135 to X.L., 
for International Leading Research
JP22K21342 to X.L., 
for Early-Career Scientists No.~JP25K18074 to T.Y.


\clearpage

\begin{figure}[h!]
\centering
\includegraphics[width=\linewidth]{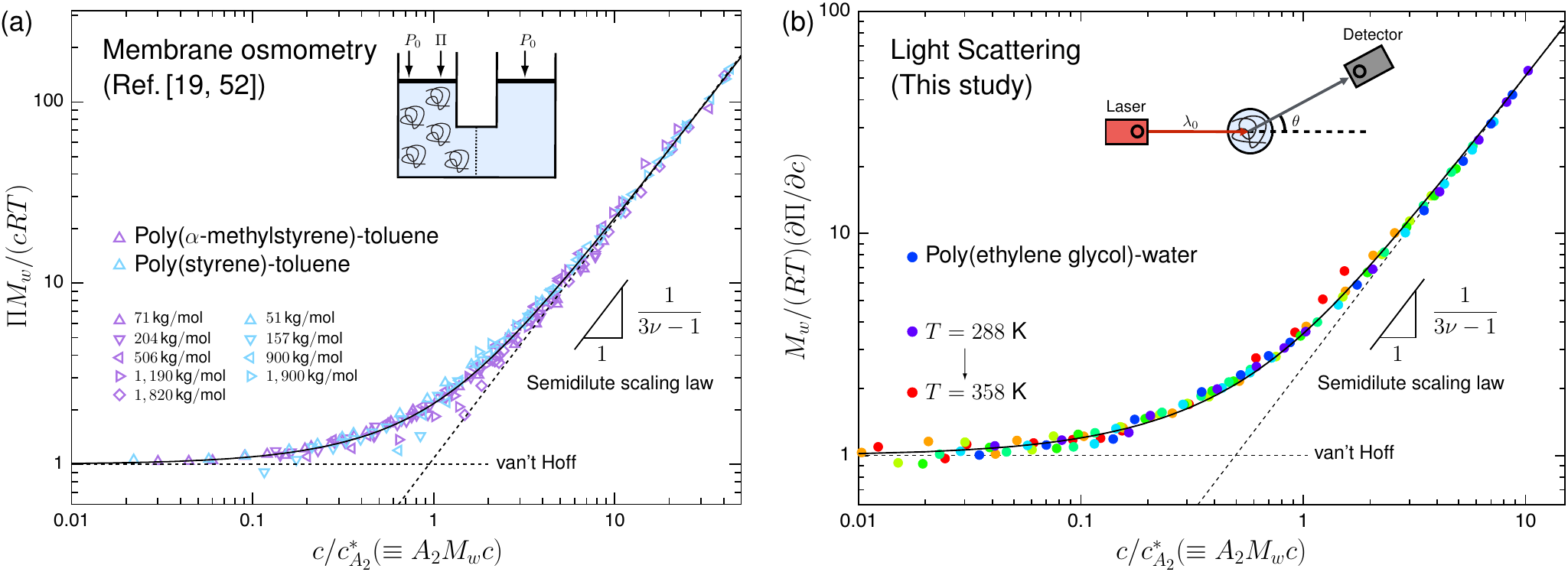}
\caption{Comparison of the universal EOS for polymer solutions in good solvents (each black solid curve) for 
(a) $\Pi$ and (b) $\partial\Pi /\partial c $.
In (a), $\Pi$ is measured by membrane osmometry (schematic) for poly($\alpha$-methylstyrene)-toluene solutions over $M_w= 71$--$1820$ kg$/$mol \cite{Noda1981_EOS} (purple triangles) and for poly(styrene)-toluene solutions over $M_w=51$--$1900$ kg$/$mol \cite{Higo1983_EOS} (light blue triangles), with $A_2$ and $M_w$ measured from light scattering to calculate the normalized quantities $\Pi M_w/(cRT)$ and $c/c^*_{A_2}\,(\equiv A_2M_wc)$.
In (b), $\partial\Pi /\partial c$ is measured by light scattering for aqueous PEG over $T=288$--$358$ K using $R_{\theta=0,f}$ (filled circles, this study).
The universal EOS is asymptotic to the van't Hoff law for $c/c^*_{A_2} \to 0$ and to the semidilute scaling law with exponent $1/(3\nu-1)$ for $c/c^*_{A_2} \gg 1$ (black dashed lines).}
\label{fig:PUEOS_EndMatter}
\end{figure}

\onecolumngrid
\vskip\baselineskip
\begin{center}
\textbf{\large End Matter}
\end{center}
\twocolumngrid
\vskip0.5\baselineskip
For neutral and flexible linear polymer solutions in good solvents, systematic osmotic pressure measurements in the 1980s \cite{Noda1981_EOS,Higo1983_EOS} established the universal EOS.
Figure~\ref{fig:PUEOS_EndMatter}(a) summarizes the osmotic pressure $\Pi$ measured by membrane osmometry (schematic) for poly($\alpha$-methylstyrene)-toluene over $M_w=71$--$1820$ kg$/$mol \cite{Noda1981_EOS} (purple triangles) and poly(styrene)-toluene over $M_w=51$--$1900$ kg$/$mol \cite{Higo1983_EOS} (light blue triangles), with the data normalized by $\Pi M_w/(cRT)$ and plotted as a function of $c/c^*_{A_2}\,(\equiv A_2M_wc)$.
The data collapse onto a single black solid curve, independent of molar mass and chemical species.
The curve crosses over from the van't Hoff law in the dilute limit ($c/c^*_{A_2} \to 0$) to the semidilute scaling law with an exponent of $1/(3\nu-1)$ in the semidilute regime ($c/c^*_{A_2}\gg 1$), as indicated by the black dashed lines.

The theoretical formulation of the universal EOS has been developed within the framework of renormalization-group theory and has undergone continuous refinement \cite{Ohta1982_EOS,cherayil1986osmotic,Merkle1993_LS}.
The functional form that quantitatively reproduces the data, however, remains unsettled.
For direct comparison with our measurements, we therefore introduce a phenomenological expression that is not based on a physical model but describes the data accurately.
To describe the crossover between the second-order virial expansion of $\Pi$ and its semidilute scaling law, corresponding to Eqs.~(\ref{eq:VE}) and (\ref{eq:scaling}), we write it as
\begin{equation}
\frac{\Pi M_w}{cRT}=\left\{(1 + c/c^*_{A_2})^p + \left[ k(c/c^*_{A_2})^{1/(3\nu-1)}\right]^p\right\}^{1/p},
\label{eq:UEOS}
\end{equation}
using two numerical fitting parameters, $k$ and $p$.
Equation~(\ref{eq:UEOS}) recovers the second-order virial expansion for the dilute regime ($c < c^*_{A_2}$)
and the semidilute scaling law for the semidilute regime ($c \gg c^*_{A_2}$).
With $\nu = 0.588$, the best fit gives $k\approx1.04$ and $p\approx2.41$, and reproduces the osmometry data [black solid curve in Fig.~\ref{fig:PUEOS_EndMatter}(a)].

Light scattering measures $\partial \Pi/\partial c$, whereas membrane osmometry measures $\Pi$, so the curve in Fig.~\ref{fig:PUEOS_EndMatter}(b) is the derivative of the curve in Fig.~\ref{fig:PUEOS_EndMatter}(a).
Figure~\ref{fig:PUEOS_EndMatter}(b) shows that curve, obtained by multiplying Eq.~(\ref{eq:UEOS}) by $c$ and differentiating, together with our measured $\partial \Pi/\partial c$ for aqueous PEG solutions over $T=288$--$358$ K from the resolved $R_{\theta=0,f}$ (filled circles).
Because the universal EOS for $\partial \Pi/\partial c$ in Fig.~\ref{fig:PUEOS_EndMatter}(b) is physically equivalent to that for $\Pi$ in Fig.~\ref{fig:PUEOS_EndMatter}(a), the agreement in the main panel of Fig.~\ref{fig:PUEOS} establishes that the resolved data follow the universal EOS.

\onecolumngrid

\end{document}


\title{Supplemental Material for: ``Mode-Resolved Light Scattering Recovers Polymer Thermodynamics in Solutions with Trace Large-Mass Scatterers''}

\author{Noriaki Mizumoto}
\thanks{These authors contributed equally: N. Mizumoto, T. Yasuda}
\affiliation{Division of Soft Matter, Graduate School of Life Science, Hokkaido University, Sapporo 060-0810, Japan.}
\author{Takashi Yasuda}
\thanks{These authors contributed equally: N. Mizumoto, T. Yasuda}
\affiliation{Faculty of Advanced Life Science, Hokkaido
University, Sapporo 001-0021, Japan.}
\author{Hiroshi Hinou}
\affiliation{Faculty of Advanced Life Science, Hokkaido
University, Sapporo 001-0021, Japan.}
\author{Xiang Li}
\email[Contact author:]{\,
x.li@sci.hokudai.ac.jp}
\affiliation{Faculty of Advanced Life Science, Hokkaido
University, Sapporo 001-0021, Japan.}

\date{\today}

\maketitle
\noindent

\setcounter{equation}{0}
\setcounter{figure}{0}
\setcounter{table}{0}
\makeatletter

\def\theequation{S\arabic{equation}}
\def\thefigure{S\arabic{figure}}
\def\thetable{S\arabic{table}}

\setcounter{figure}{0}
\renewcommand{\thesection}{S\arabic{section}}
\renewcommand{\thefigure}{S\arabic{figure}}
\renewcommand{\thetable}{S\arabic{table}}
\renewcommand{\bibnumfmt}[1]{[S#1]}
\renewcommand{\citenumfont}[1]{S#1}

\section{Derivation of the equations for mode-resolved analytical framework}

We present the derivation of Eqs.~(1) and (2) in the main text, accounting for solvent scattering contributions and applying the Siegert relation.
When the dynamics of the solvent, fast mode, and slow mode are independent, the field correlation function $\tilde {g}^{(1)}(\tau)$ is followed as
\begin{equation}
\tilde{g}^{(1)}(\tau)
=\tilde{a}_{\mathrm{solv}}\exp{(-\Gamma_{\mathrm{solv}}\tau)}
+\tilde{a}_{f}\exp{(-\Gamma_{f}\tau)}
+\tilde{a}_{s}\exp{(-\Gamma_{s}\tau)},
\label{eq:g1_solvent}
\end{equation}
where $\Gamma_{\mathrm{solv}}$, $\Gamma_{f}$, and $\Gamma_{s}$ are the relaxation rates of the solvent molecule, fast mode, and slow mode, respectively; $\tilde{a}_{\mathrm{solv}}$, $\tilde{a}_{f}$ and $\tilde{a}_{s}$ are their fractional intensity contributions $(\tilde{a}_{\mathrm{solv}}+\tilde{a}_{f}+\tilde{a}_{s}=1)$; and $\tau$ is the lag time.
The $\tilde{g}^{(1)}(\tau)$ is related to the intensity correlation function $g^{(2)}(\tau)$ through the Siegert relation:
\begin{equation}
g^{(2)}(\tau) = 1 + \beta f(I) |\tilde{g}^{(1)}(\tau)|^2,
\label{eq:Siegert}
\end{equation}
where $\beta$ is the intrinsic optical coherence factor of the instrument, and $f(I)=[(1-\tau_dI)(1-b/I)]^2$ is the detector correction function determined by the dead time of the detector $\tau_d$, total detector count rate $I$, and the detector noise $b$ (see Supporting Information in Ref.~\cite{Rochas2014_beta} for details).
The total Rayleigh ratio of the measured solution $R_{\theta,\mathrm{solution}} = R_{\theta,\mathrm{solv}} + R_{\theta}$ consists of contributions from solvent scattering $R_{\theta,\mathrm{solv}}$ and overall solute scattering $R_\theta= R_{\theta,f} + R_{\theta,s}$, each of which can be further resolved into mode-specific contributions \cite{Raspaud1994_theory}:
\begin{equation}
R_{\theta,\mathrm{solv}} = \tilde{a}_\mathrm{solv} R_{\theta,\mathrm{solution}},\ 
R_{\theta,f} = \tilde{a}_f R_{\theta,\mathrm{solution}},\ 
R_{\theta,s} = \tilde{a}_s R_{\theta,\mathrm{solution}}.
\label{eq:resolved_total}
\end{equation}

When solvent relaxation fully decays ($\tau \gg \Gamma_\mathrm{solv}^{-1}$), Eq.~(\ref{eq:g1_solvent}) reduces to 
$\tilde{g}^{(1)}(\tau)=\tilde{a}_{f}\exp{(-\Gamma_{f}\tau)}
+\tilde{a}_{s}\exp{(-\Gamma_{s}\tau)}$.
Using $a_f \equiv \tilde{a}_f / (\tilde{a}_f + \tilde{a}_s)$ and $a_s \equiv \tilde{a}_s / (\tilde{a}_f + \tilde{a}_s)$ with $\tilde{a}_f + \tilde{a}_s = 1 - \tilde{a}_\mathrm{solv}$, $\tilde{g}^{(1)}(\tau)$ is expressed as
\begin{equation}
\tilde{g}^{(1)}(\tau) = (1 - \tilde{a}_\mathrm{solv})[a_{f}\exp{(-\Gamma_{f}\tau)}
+a_{s}\exp{(-\Gamma_{s}\tau)}].
\label{eq:g1_withoutsolvent}
\end{equation}
Substituting Eq.~(\ref{eq:g1_withoutsolvent}) into Eq.~(\ref{eq:Siegert}), we have
\begin{equation}
g^{(2)}(\tau) 
= 1 + \beta f(I) (1-\tilde{a}_{\mathrm{solv}})^2
\left|
a_f\exp({-\Gamma_f\tau})
+a_s\exp({-\Gamma_s\tau})
\right|^2.
\label{eq:Siegert_beta_eff}
\end{equation}
Defining $g^{(1)}(\tau) = a_f\exp({-\Gamma_f\tau})+a_s\exp({-\Gamma_s\tau})$ for the case without the solvent scattering contribution ($a_f + a_s = 1$), 
Eq.~(\ref{eq:Siegert_beta_eff}) is expressed as
$g^{(2)}(\tau)=1 + \beta_{\mathrm{eff}}|g^{(1)}(\tau)|^2,$
where $\beta_{\mathrm{eff}} \equiv \beta f(I) (1-\tilde{a}_{\mathrm{solv}})^2$ is the effective coherence factor described in the main text.
The defined $g^{(1)}(\tau)\,[= \tilde{g}^{(1)}(\tau)/(1 - \tilde{a}_\mathrm{solv})]$ corresponds to Eq.~(1) in the main text.
Because $R_\theta = (\tilde{a}_f + \tilde{a}_s)R_{\theta,\mathrm{solution}}$, Eq.~(\ref{eq:resolved_total}) reduces to $R_{\theta,f} = a_f R_\theta$ and $R_{\theta,s} = a_s R_\theta$, which corresponds to Eq.~(2) in the main text.

\begin{figure}[t!]
\centering
\includegraphics[width=0.6\linewidth]{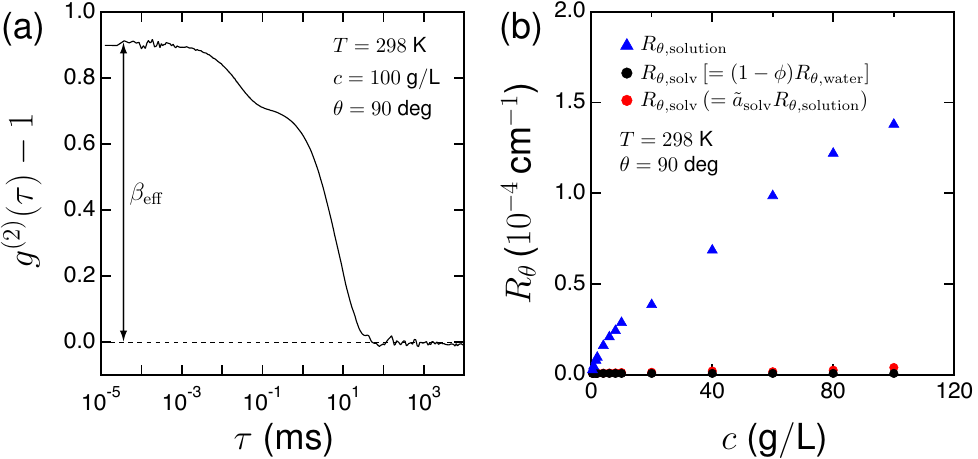}
\caption{(a) Typical $g^{(2)}(\tau)-1$ profile for $c=100$ g$/$L at $T=298$ K  and $\theta=90^{\circ}$, with the least-square fit to the baseline with a constant (black dashed line).
(b) The $c$ dependence of $R_{\theta,\mathrm{solution}}$ (blue triangles), $R_{\theta,\mathrm{solv}} = (1-\phi)R_{\theta,\mathrm{water}}$ (black circles), and $R_{\theta,\mathrm{solv}} = \tilde{a}_\mathrm{solv} R_{\theta,\mathrm{solution}}$ (red circles) at $T=298$ K and $\theta=90^{\circ}$.
}
\label{fig:derivation}
\end{figure}

As a specific example, we illustrate the treatment of the solvent scattering contribution in this study.
Figure~\ref{fig:derivation}(a) shows the typical $g^{(2)}(\tau)-1$ profile for aqueous PEG solutions measured for $c=100$ g$/$L at $T=298$ K and $\theta=90^{\circ}$, where $c$ is the polymer mass concentration (polymer weight per solute volume), $T$ is the absolute temperature, and $\theta$ is the scattering angle.
The baseline in $g^{(2)}(\tau) - 1$ profile (black dashed line) was sufficiently small to be neglected.
Because the solvent relaxation had sufficiently decayed over the measured lag-time range, $g^{(1)}(\tau)$ was calculated by taking the initial amplitude of $g^{(2)}(\tau) - 1$ in the measured lag-time range as $\beta_\mathrm{eff}$, and therefore we applied Eqs.~(1) and (2) in the main text for the mode-resolved analysis.
To obtain $R_\theta \,(= R_{\theta,\mathrm{solution}} - R_{\theta,\mathrm{solv}})$ used in Eq.~(2) in the main text, we measured the Rayleigh ratio of pure water $R_{\theta,\mathrm{water}}$ and calculating $R_{\theta,\mathrm{solv}} = (1 - \phi)R_{\theta,\mathrm{water}}$, where $\phi$ is the polymer volume fraction.
Notably, Eq.~(\ref{eq:resolved_total}) was not used for the calculation of $R_{\theta,\mathrm{solv}}$, as determining $\tilde{a}_\mathrm{solv}$ requires precise measurements of $\beta_\mathrm{eff}$ and $\beta f(I)$.
Figure~\ref{fig:derivation} (b) shows the $c$ dependence of $R_{\theta,\mathrm{solution}}$ (blue triangles) and the comparison of the $c$ dependence of $R_{\theta,\mathrm{solv}} = (1-\phi)R_{\theta,\mathrm{water}}$ (black circles) and $R_{\theta,\mathrm{solv}} = \tilde{a}_\mathrm{solv} R_{\theta,\mathrm{solution}}$ (red circles) measured at $T=298$ K and $\theta=90^{\circ}$.
Here, we assumed the constant $\beta f(I) \approx 0.97$ to determine $\tilde{a}_\mathrm{solv}$.
Both $R_{\theta,\mathrm{solv}} = (1-\phi)R_{\theta,\mathrm{water}}$ and $R_{\theta,\mathrm{solv}} = \tilde{a}_\mathrm{solv} R_{\theta,\mathrm{solution}}$ are much smaller compared to $R_{\theta,\mathrm{solution}}$, with $R_{\theta,\mathrm{solv}} = \tilde{a}_\mathrm{solv} R_{\theta,\mathrm{solution}}$ being close to $R_{\theta,\mathrm{solv}} = (1-\phi)R_{\theta,\mathrm{water}}$.
However, $R_{\theta,\mathrm{solv}} = \tilde{a}_\mathrm{solv} R_{\theta,\mathrm{solution}}$ exhibits slight concentration dependence at high concentrations, presumably arising from uncertainties in $\beta_\mathrm{eff}$ and $\beta f(I)$.

\clearpage

\section{Determination of Optical Constant from Refractive Index}

We determine the optical constant $K$ at each temperature investigated.
The optical constant is defined as \cite{zhang2009_K}
\begin{equation}
K\equiv \frac{\pi^2}{N_A\lambda_0^4}
\left(
\frac{\partial \bar{\epsilon}_r}{\partial \tilde{c}}
\right)^2,
\label{eq:K_def}
\end{equation}
where $N_A$ is the Avogadro constant; $\lambda_0$ is the incident wavelength in vacuum; $\bar{\varepsilon}_r$ is the average relative dielectric constant of the sample; and $\tilde{c}$ is the polymer mass concentration (polymer weight per solution volume).
Because $\bar{\epsilon}_r$ is related to the refractive index of the sample $n$ by $\bar{\epsilon}_r = n^2$, Eq.~(\ref{eq:K_def}) reduces to 
\begin{equation}
K  = \frac{\pi^2}{N_A\lambda_0^4}
\left[
\frac{d(n^2)}{d\tilde{c}}
\right]^2.
\label{eq:K}
\end{equation}
Since $d(n^2)/d\tilde{c} =2n(dn/d\tilde{c})$, Eq.~(\ref{eq:K}) further reduces to 
\begin{equation}
K = \frac{4\pi^2n^2}{N_A\lambda_0^4}
\left(
\frac{dn}{d\tilde{c}}
\right)^2.
\label{eq:K_c0}
\end{equation}
When $d(n^2)/d \tilde{c}$ is constant, the product $n(dn/d\tilde{c})$ is essentially independent of $\tilde{c}$, although $n$ and $dn/d\tilde{c}$ individually depend on $\tilde{c}$.
Under this condition, $K$ can be determined experimentally from the following equation:
\begin{equation}
K = \frac{4\pi^2n_s^2}{N_A\lambda_0^4}
\left(
\left.\frac{dn}{d\tilde{c}}\right|_{\tilde{c}=0}
\right)^2.
\label{eq:K_c0}
\end{equation}

\begin{figure}[h!]
\centering
\includegraphics[width=1\linewidth]{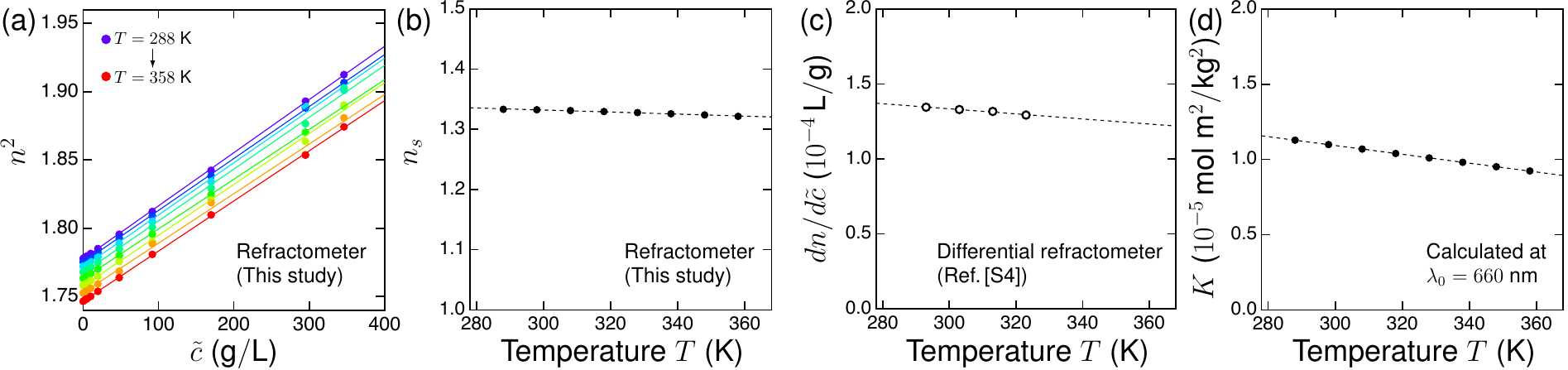}
\caption{Temperature dependence of refractive index $n$ in the linear polymer (PEG) solutions and pure solvent.
(a) The $\tilde{c}$ dependence of $n^2$ over $T=288$--$358$ K.
Each solid line represents the least-square fit to the data with a linear function. 
(b) Temperature dependence of refractive index in aqueous solvents $n_s$ over $T=288$--$358$ K.
The dashed line represents the least-square fit to the data with a linear function.
(c) Temperature dependence of $dn/d\tilde{c}$ measured by differential refractometer reported in Ref.~\cite{William1983_SLSPEOTemp20to90}.
The dashed line represents the least-square fit to the data with a linear function.
(d) Temperature dependence of $K$ at $\lambda_0=660$ nm.
The dashed line represents the least-square fit to the data with a linear function.
}
\label{fig:RI}
\end{figure}

We confirmed that $d(n^2)/d\tilde{c}$ is approximately constant over the concentration range investigated, implying $n(dn/d\tilde{c})$ is also constant.
Figure~\ref{fig:RI}(a) shows $\tilde{c}$ dependence of the squared refractive index $n^2$ for aqueous PEG solutions and $n_s^2$ for pure aqueous solvent over $T=288$--$358$ K, measured at $\lambda_0 = 589$ nm with a refractometer (Abbemat 650, Anton Paar, Austria).
At each $T$, $n^2 - n_s^2$ is approximately proportional to $\tilde{c}$ (each solid line).
As $T$ increases, both $n^2$ and $n_s^2$ decrease.
The $T$ dependence of $n_s$ is shown in Fig.~\ref{fig:RI}(b).

Using measured $n_s$ together with the literature value of $dn/d\tilde{c}$ shown in Fig.~\ref{fig:RI}(c), we determined $K$ for aqueous PEG solutions at each temperature investigated in this study.
The previously reported $dn/d\tilde{c}$ used in the calculation slightly decreases with increasing temperature (dashed line), obtained from measurements for dilute aqueous PEG solutions over $T=293$--$323$ K at $\lambda_0=633$ nm with a differential refractometer \cite{William1983_SLSPEOTemp20to90}.
Because $dn/d\tilde{c}$ is approximately independent of the wavelength over the visible wavelength range \cite{William1983_SLSPEOTemp20to90}, we use $n_s$ shown in Fig.~\ref{fig:RI}(b) and the interpolation (or extrapolation) of $dn/d\tilde{c}$ shown in Fig.~\ref{fig:RI}(c) (dashed line) for the evaluation of $K$ at $\lambda_0=660$ nm.
Figure~\ref{fig:RI}(d) shows the temperature dependence of $K$ evaluated at $\lambda_0=660$ nm, which indicates that $K$ decreases with increasing temperature (dashed line).
The obtained $K$ at each $T$ was used for the analysis of $\partial\Pi / \partial c$ in Eq.~(3) in the main text.

\newpage

\section{MALDI-TOF MS experiment}

Sample was analyzed by matrix-assisted laser desorption/ionization time-of-flight mass spectrometry (MALDI-TOF MS) using a Ultraflex III MALDI-TOF/TOF instrument (Bruker, Bremen, Germany) equipped with a $200$ Hz smartbeam Nd:YAG UV laser (wavelength of $355$ nm) in the positive linear mode.
A $0.5$ $\mathrm{\mu}$L aliquot of the PEG solution ($10$ g$/$L in water) and a $1.0$ $\mathrm{\mu}$L aliquot of 2,5-dihydroxybenzoic acid [$10$ g$/$L in an equal-volume mixture of acetonitrile and water] ware mixed on a MALDI plate (STA $\mathrm{\mu}$Focus plate, $24 \times 16$, $700$ $\mathrm{\mu}$m) (Hudson Surface Technology, Inc., Closter, NJ, USA), and air-dried before measurement.

\begin{figure}[h!]
\centering
\includegraphics[width=0.3\linewidth]{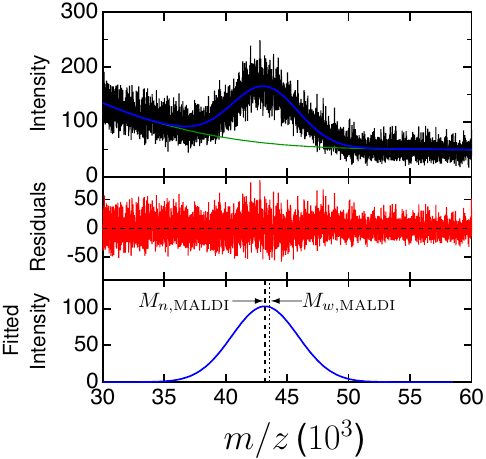}
\caption{MALDI-TOF MS spectrum and Gaussian peak fitting analysis of the linear methoxy-PEG-OH (nominal $M_{n,\mathrm{nom}} \approx 40$ kg$/$mol, $\mathrm{PDI}_{\mathrm{nom}} \approx 1.2$) used in this study.
Top panel: experimental spectrum (black), fitted baseline (green), and total fit (blue curve).
Middle panel: fit residuals (red).
Bottom panel: fit curve of Gaussian peak (blue) with indicated $M_{n,\mathrm{MALDI}}\approx 43.2$ kg$/$mol and $M_{w,\mathrm{MALDI}} \approx 43.5$ kg$/$mol, yielding $\mathrm{PDI}_{\mathrm{MALDI}} \approx 1.01$.}
\label{fig:MALDI}
\end{figure}

To determine the number-average molar mass $M_{n,\mathrm{MALDI}}$, the weight-average molar mass $M_{w,\mathrm{MALDI}}$, and the polydispersity $\mathrm{PDI}_{\mathrm{MALDI}}$, the MALDI-TOF MS spectrum was analyzed by baseline subtraction followed by a single-Gaussian fit (Fig.~\ref{fig:MALDI}).
Assuming that the polymer molecules were ionized predominantly as singly charged species (charge number $z$ = 1), we treated the $m/z$ scale, the standard mass-to-charge ratio of mass spectrometry, directly as the molar mass.
The fit gave the peak center $M_0 \approx 43.2$ kg$/$mol and the standard deviation $\sigma \approx 3.8$ kg$/$mol.
The $M_{n,\mathrm{MALDI}}$ and $M_{w,\mathrm{MALDI}}$ were calculated from the fitted parameters ($M_{n,\mathrm{MALDI}}=M_0,\, M_{w,\mathrm{MALDI}} = M_0 + \sigma^2/M_0$), yielding $M_{n,\mathrm{MALDI}}\approx 43.2$ kg$/$mol, $M_{w,\mathrm{MALDI}}\approx 43.5$ kg$/$mol, and $\mathrm{PDI}_{\mathrm{MALDI}}\approx 1.01$.

\newpage

\section{Temperature dependence of $R_{\theta,\mathrm{water}}$}

\begin{figure}[h!]
\centering
\includegraphics[width=0.3\linewidth]{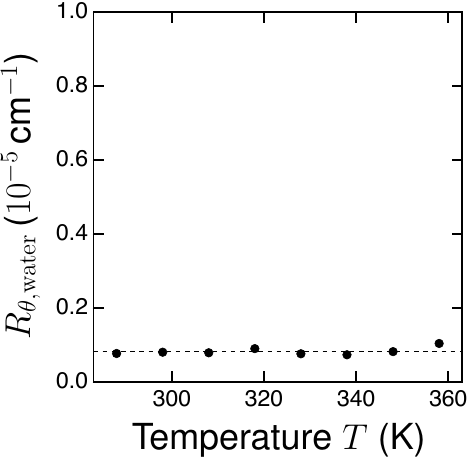}
\caption{The $T$ dependence of $R_{\theta,\mathrm{water}}$, being approximately constant (black dashed line).
}
\label{fig:SolventR}
\end{figure}

We show the $T$ dependence of measured Rayleigh ratio of aqueous solvent $R_{\theta,\mathrm{water}}$ over $T=288$--$358$ K in Fig.~\ref{fig:SolventR}.
The $R_{\theta,\mathrm{water}}$ is approximately independent of $T$ (black dashed line).
In this study, we use $R_{\theta,\mathrm{water}}$ at each $T$ to calculate $R_\theta$ for each measured $R_{\theta,\mathrm{solution}}$.

\section{All $g^{(1)}(\tau)$ profiles over varying scattering angle and temperature}

\begin{figure}[h!]
\centering
\includegraphics[width=0.8\linewidth]{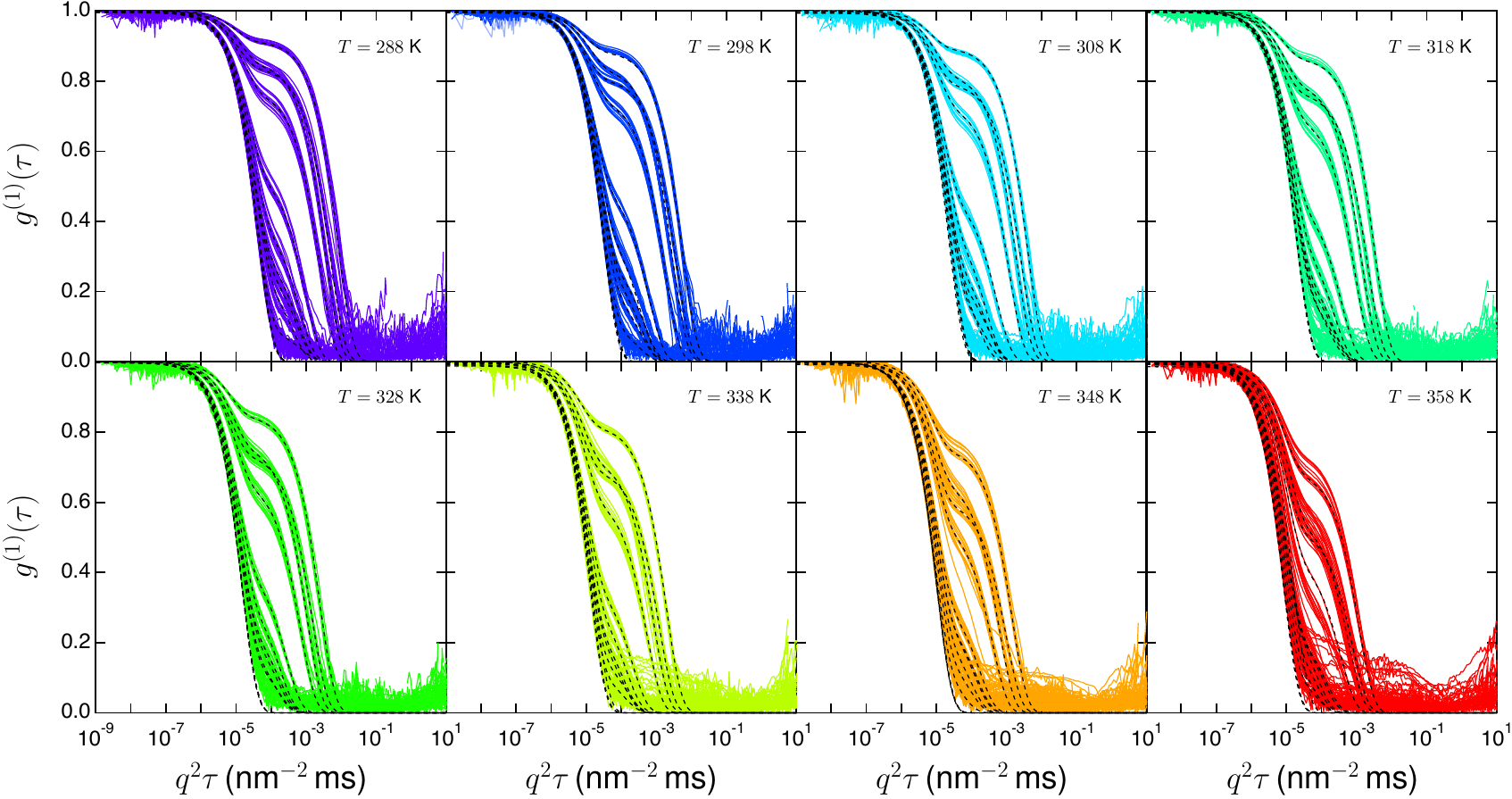}
\caption{Summary of the $q^2 \tau$ dependence of $g^{(1)}(\tau)$ over $T=288$--$358$ K for $c=0.4$--$100$ g$/$L in aqueous polymer (PEG) solutions.
Each black dashed curve represents the least-squares fit to $g^{(1)}(\tau)$ at $\theta = 90^{\circ}$ with Eq.~(1) in the main text for each $c$.}
\label{fig:g1_Temp}
\end{figure}

We summarize all data on the $q^2\tau$ dependence of $g^{(1)}(\tau)$ over $\theta = 50^{\circ}$--$150^{\circ}$ for $c = 0.4$--$100$ g$/$L over $T = 288$--$358$ K in Fig.~\ref{fig:g1_Temp}, where $q$ is the scattering vector magnitude.
Here, $g^{(1)}(\tau)$ at $\theta = 90^{\circ}$ and $T = 298$ K is presented in Fig.~2(a) in the main text.
At all $T$, $g^{(1)}(\tau)$ exhibits the pronounced slow mode as the increase in $c$.
All $g^{(1)}(\tau)$ are well described by Eq.~(1) in the main text (black solid curves), yielding $\Gamma_f$, $\Gamma_s$, $a_f$ ,and $a_s$.

\newpage

\section{ANALYSIS OF OBTAINED $\Gamma_f$, $\Gamma_s$, $a_f$, and $a_s$}

\begin{figure}[h!]
\centering
\includegraphics[width=1\linewidth]{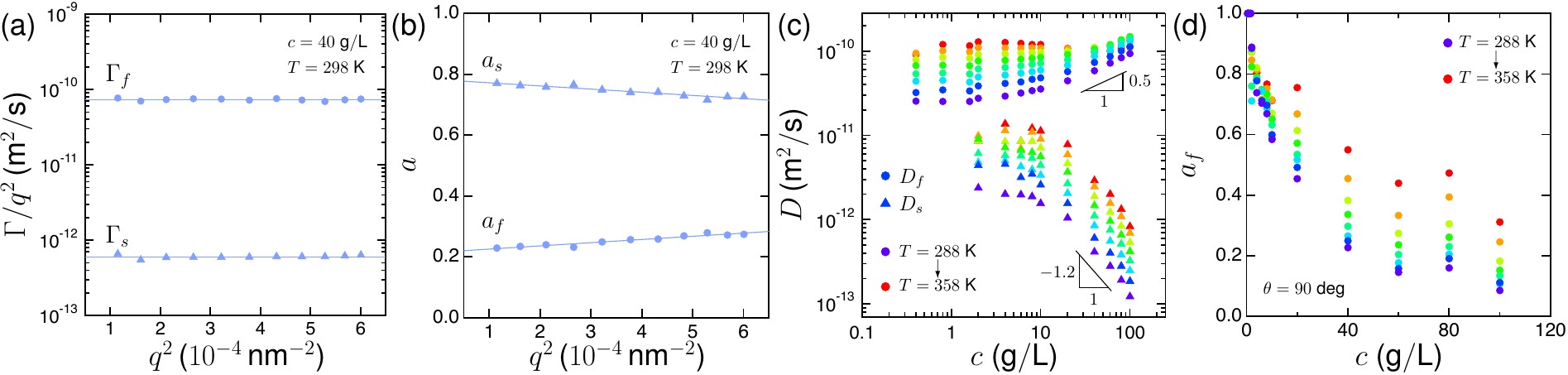}
\caption{Analysis of obtained $\Gamma_f$, $\Gamma_s$, $a_f$, and $a_s$.
(a) The representative $q^2$ dependence of $\Gamma_f/q^2$ (circles) and $\Gamma_s/q^2$ (triangles) for $c = 40$ g$/$L at $T = 298$ K. 
Each line represents the least-squares fit to the data with a constant.
(b) The representative $q^2$ dependence of $a_f$ (circles) and $a_s\,(=1-a_f)$ (triangles) for $c = 40$ g$/$L at $T = 298$ K. 
Each line represents the least-squares fit to the data with a linear function.
(c) The $c$ dependence of $D_f$ (circles) and $D_s$ (triangles) over $T = 288$--$358$ K.
Here, each of $D_f$ and $D_s$ is determined by the average of $\Gamma_f/q^2$ and $\Gamma_s/q^2$ over $\theta = 50^{\circ}$--$150^{\circ}$, respectively, which is independent of $q^2$.
(d) The $c$ dependence of $a_f$ at $\theta = 90^{\circ}$ over $T = 288$--$358$ K. 
Note that $a_f \approx 1$ for $c = 0.4$--$1.6$ g$/$L.
}
\label{fig:Temp_Df_Ds_af}
\end{figure}

We characterize the fast and slow modes by analyzing the obtained $\Gamma_f$, $\Gamma_s$, $a_f$, and $a_s$ from $g^{(1)}(\tau)$ in Fig.~\ref{fig:g1_Temp}.
Figure~\ref{fig:Temp_Df_Ds_af}(a) shows the representative $q^2$ dependence of $\Gamma_f/q^2$ (circles) and $\Gamma_s/q^2$ (triangles) for $c=40$ g$/$L at $T=298$ K (see Sec.~S7 for other $c$ and $T$).
Both are independent of $q^2$ (lines) to confirm $\Gamma_f \propto q^2$ and $\Gamma_s \propto q^2$, indicating that both modes arise from Brownian motion \cite{Hulst1957_LS,BerneandPecora1976_LS}.
The observed $q^2$ independence is consistent with the previous reports in the fast \cite{Hiroi2014_diffusion} and slow  \cite{Alami1996_slowmode} modes.
Figure~\ref{fig:Temp_Df_Ds_af}(b) shows the representative $q^2$ dependence of $a_f$ (circles) and $a_s$ (triangles) for $c=40$ g$/$L at $T=298$ K (see Sec.~S8 for other $c$ and $T$).
Both are dependent on $q^2$ (lines), with $a_f$ increasing and $a_s$ decreasing as $q^2$ increases.
These $q^2$ dependence of $a_f\,(=R_{\theta,f}/R_\theta)$ and $a_s\,(=R_{\theta,s}/R_\theta)$ are arising from the $q$ dependence of $R_\theta$ and its constituent of $R_{\theta,f}$ and $R_{\theta,s}$, as shown in Fig.~3(a) in the main text.

We summarize the $c$ dependence of the diffusion coefficients of the fast mode $D_f$ (circles) and the slow mode $D_s$ (triangles) over $T=288$--$358$ K in Fig.~\ref{fig:Temp_Df_Ds_af}(c).
The $D_f$ and $D_s$ are determined from the average of $\Gamma_f/q^2$ and $\Gamma_s/q^2$ measured over $\theta = 50^{\circ}$--$150^{\circ}$ [lines in Fig.~\ref{fig:Temp_Df_Ds_af}(a)].
The $D_f$ ranges from $10^{-10}$--$10^{-11}$ m$^2$$/$s, which is consistent with the typical Brownian diffusion of solvated polymer chains \cite{Hiroi2014_diffusion,Fujiyabu2019_diffusion}.
For the dilute limit ($c \to 0$), $D_f$ is asymptotic to a constant to approach the self-diffusion coefficient, which corresponds to the hydrodynamic radius $R_h$ in the Stokes-Einstein relation \cite{Einstein1905_SE}.
As $c$ increases, $D_f$ increases in the intermediate regime ($D \sim c^{0.5}$ as a guide line to the measured data) and is expected to asymptotically approach $D\sim c^{\nu/(3\nu-1)} \sim c^{0.77}$ for the semidilute limit ($c \gg c^*_{A_2}$), where $\nu \approx 0.588$ \cite{Gennes1979_polymer}.
On the other hand, $D_s$ ranges from $10^{-11}$--$10^{-13}$ m$^2$$/$s, which is consistent with the order reported in previous reports 
\cite{Brown1984_slowmode,Brown1985_slowmode,Alami1996_slowmode}, and exhibits a significant decrease as $c$ increases ($D \sim c^{-1.2}$ as a guide line to the measured data).
This decrease in $D_s$ closely resembles the diffusion coefficient of gold nanoparticles mixed in the semidilute polymer solutions \cite{Koziol2019_slowmode}, indicating the suppression of the diffusion due to the increase in the environment polymer concentration.
Notably, both $D_f$ and $D_s$ increase as $T$ increases, which is attributed to the $T$ dependence of the viscosity $\eta_0$ of solvents ($D\propto \eta_0^{-1}$).

We summarize the $c$ dependence of $a_f\,(=1-a_s)$ at the representative angle $\theta = 90^{\circ}$ over $T=288$--$358$ K in Fig.~\ref{fig:Temp_Df_Ds_af}(d).
The $a_f$ generally decreases as $c$ increases, indicating that the fraction of the slow mode is significant in higher concentrations to dominate the Rayleigh ratio.
In particular, $a_f\approx1$ for $c=0.4$--$1.6$ g$/$L, indicating that the slow mode is not observed in sufficiently dilute concentrations.
Additionally, $a_f$ generally increases as $T$ increases, which arises from the $T$ dependence of $R_{\theta,f}$ and $R_{\theta,s}$.

\newpage

\section{All data on the $q^2$ dependence of $\Gamma_f/q^2$ and $\Gamma_s/q^2$}

\begin{figure}[h!]
\centering
\includegraphics[width=0.8\linewidth]{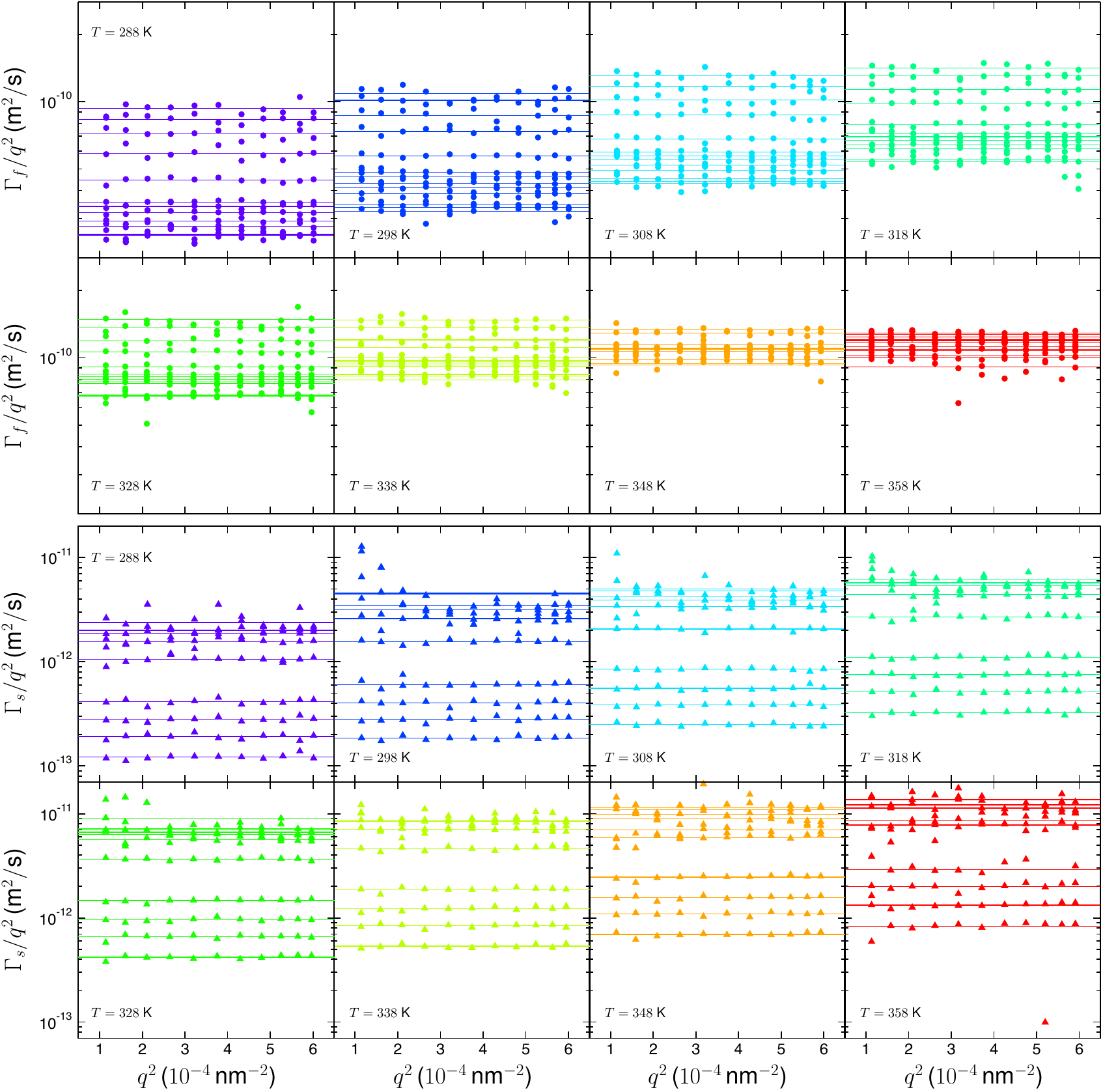}
\caption{The summary of the $q^2$ dependence of $\Gamma_f/q^2$ (upper panel) and $\Gamma_s/q^2$ (lower panel) over $T=288$--$358$ K.
Each line represents the least-squares fit to the data with a constant, excluding the outlier data.}
\label{fig:Gamma_f_s_q2}
\end{figure}

We summarize all data on the $q^2$ dependence of $\Gamma_f/q^2$ (upper panel) and $\Gamma_s/q^2$ (lower panel) over $T=288$--$358$ K in Fig.~\ref{fig:Gamma_f_s_q2}.
Here, $\Gamma_f/q^2$ and $\Gamma_s/q^2$ at $T = 298$ K for $c=40$ g$/$L are presented in Fig.~\ref{fig:Temp_Df_Ds_af}(a).
At all $T$, both $\Gamma_f/q^2$ and $\Gamma_s/q^2$ are approximately independent of $q^2$ except for the outlier data (lines), confirming $\Gamma_f \propto q^2$ and $\Gamma_s \propto q^2$ as the mode arising from the Brownian motion \cite{Hulst1957_LS,BerneandPecora1976_LS}.

\newpage

\section{All data on the $q^2$ dependence of $a_f$}

\begin{figure}[h!]
\centering
\includegraphics[width=0.8\linewidth]{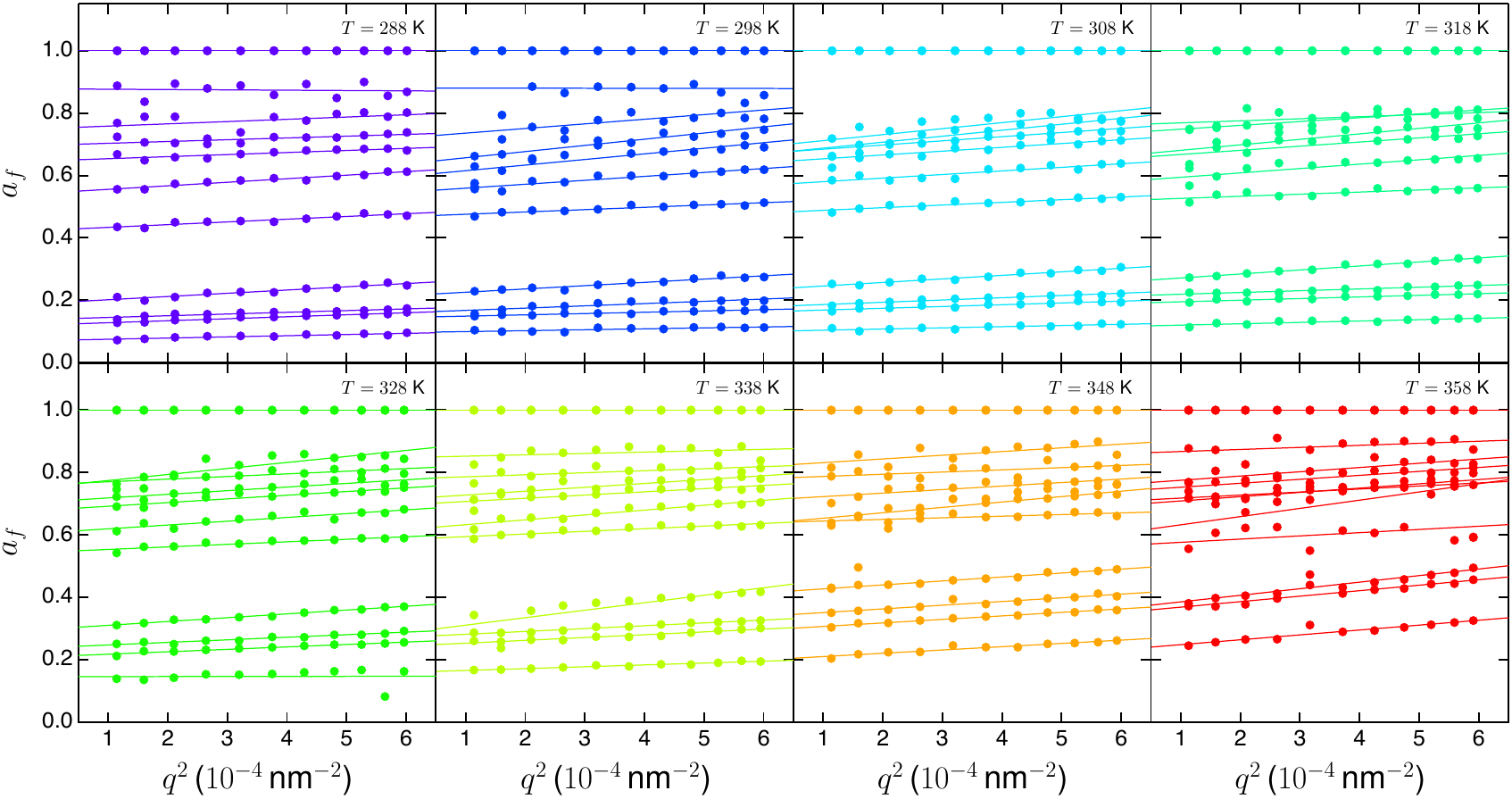}
\caption{The summary of the $q^2$ dependence of $a_f$ over $T=288$--$358$ K.}
\label{fig:a_f_q2}
\end{figure}

We summarize all data on the $q^2$ dependence of $a_f\,(=1-a_s)$ over $T=288$--$358$ K in Fig.~\ref{fig:a_f_q2}.
Here, $a_f$ at $T = 298$ K for $c=40$ g$/$L is presented in Fig.~\ref{fig:Temp_Df_Ds_af}(b).
In addition, $a_f$ at $\theta = 90^\circ$ over $T=288$--$358$ K are presented in Fig.~\ref{fig:Temp_Df_Ds_af}(d). 
We use each $a_f$ to calculate $R_{\theta,f}$ and $R_{\theta,s}$ in Fig.~\ref{fig:R_f_s_q2}, using Eq.~(2) in the main text.

\section{All data of resolving $R_{\theta,f}$ from $R_\theta$}

\begin{figure*}[h!]
\centering
\includegraphics[width=0.8\linewidth]{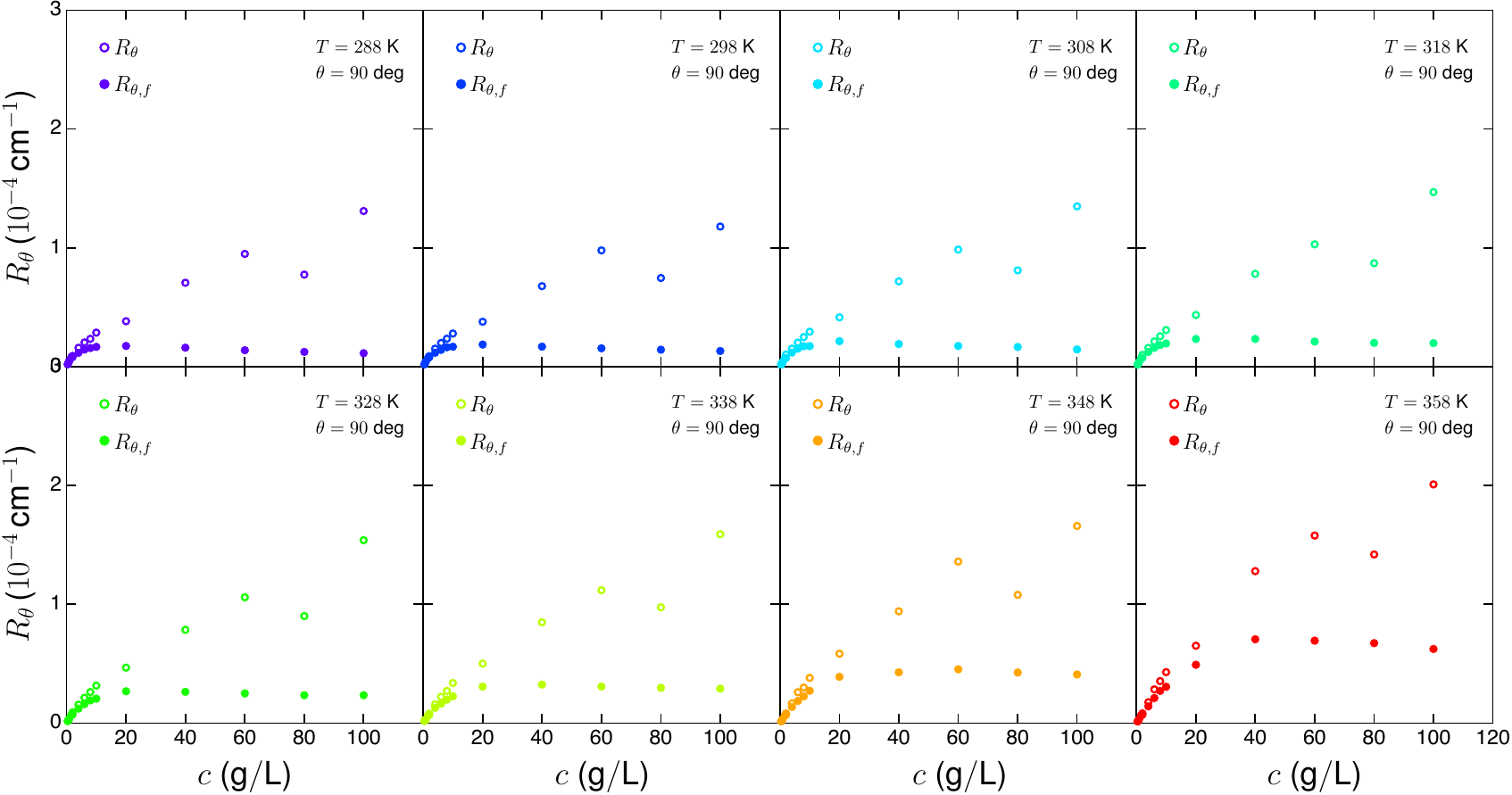}
\caption{The $c$ dependence of $R_{\theta,f}$ (filled circles) resolved from $R_{\theta}$ (open circles) at $\theta=90^{\circ}$ over $T=288$--$358$ K, using Eq.~(2) in the main text with each $a_f$ evaluated in Fig.~\ref{fig:a_f_q2}.}
\label{fig:R_f_tot_c}
\end{figure*}

We summarize all data on the $c$ dependence of $R_{\theta,f}$ (filled circles) resolved from $R_{\theta}$ (open circles) at $\theta=90^{\circ}$ over $T=288$--$358$ K in Fig.~\ref{fig:R_f_tot_c}, using Eq.~(2) in the main text with each $a_f$ evaluated in Fig.~\ref{fig:a_f_q2}.
At all $\theta$ and $T$, $R_{\theta,f}$ is successfully resolved to provide the $q$ dependence of $R_{\theta,f}$, as represented in Fig.~3(a) in the main text.

\newpage

\section{Reproducibility in $g^{(1)}(\tau)$ profiles on the same sample}
\begin{figure}[h!]
\centering
\includegraphics[width=0.5\linewidth]{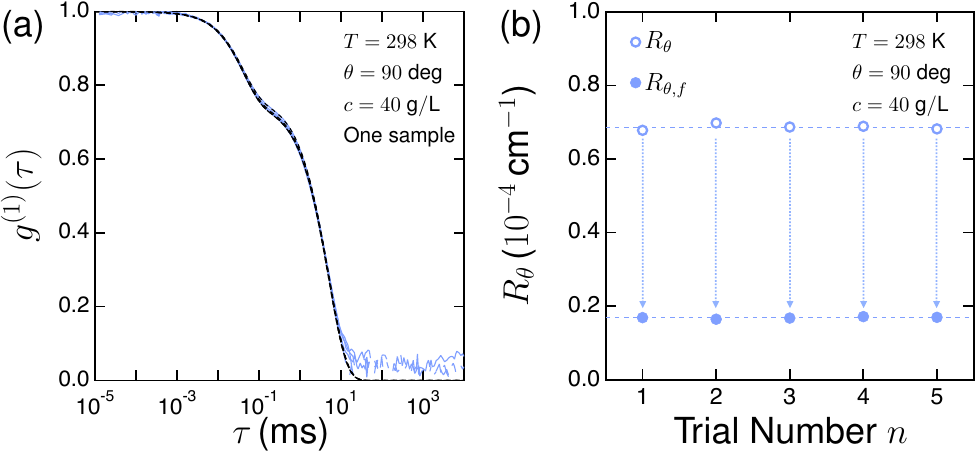}
\caption{(a) The $g^{(1)}(\tau)$ profile measured for five times on the same sample for $c=40$ g$/$L at $T=298$ K and $\theta=90^{\circ{}}$.
Each black dashed curve represents the least-squares fit to $g^{(1)}(\tau)$ with Eq.~(1) in the main text.
(b) Resolving $R_{\theta,f}$ (filled circles) from $R_{\theta}$ (open circles) for each trial number on the same sample, using Eq.~(2) in the main text with $a_f$ evaluated in (a).
The blue dashed lines represent the least-squares fit to the resolved $R_{\theta,f}$ and $R_{\theta}$ with a constant.
}
\label{fig:Rayleigh_trial}
\end{figure}

We verify the reproducibility of both $R_\theta$ and $R_{\theta,f}$ when measured multiple times on the same sample, as generally expected.
Figure~\ref{fig:Rayleigh_trial}(a) shows the $g^{(1)}(\tau)$ profile for $c=40$ g$/$L at $T=298$ K and $\theta=90^{\circ}$, measured five times on the same sample.
Using $a_f$ obtained from the fit by Eq.~(1) in the main text (each black dashed curve), Fig.~\ref{fig:Rayleigh_trial}(b) shows $R_{\theta,f}$ (filled circles) resolved from $R_{\theta}$ (open circles) for each trial number on the same sample.
Both $R_{\theta,f}$ and $R_{\theta}$ exhibit a constant (blue dashed lines), verifying robust reproducibility of light scattering measurements.

\section{Non-Reproducibility of slow mode in $g^{(1)}(\tau)$ profiles on distinct samples}

\begin{figure}[h!]
\centering
\includegraphics[width=1\linewidth]{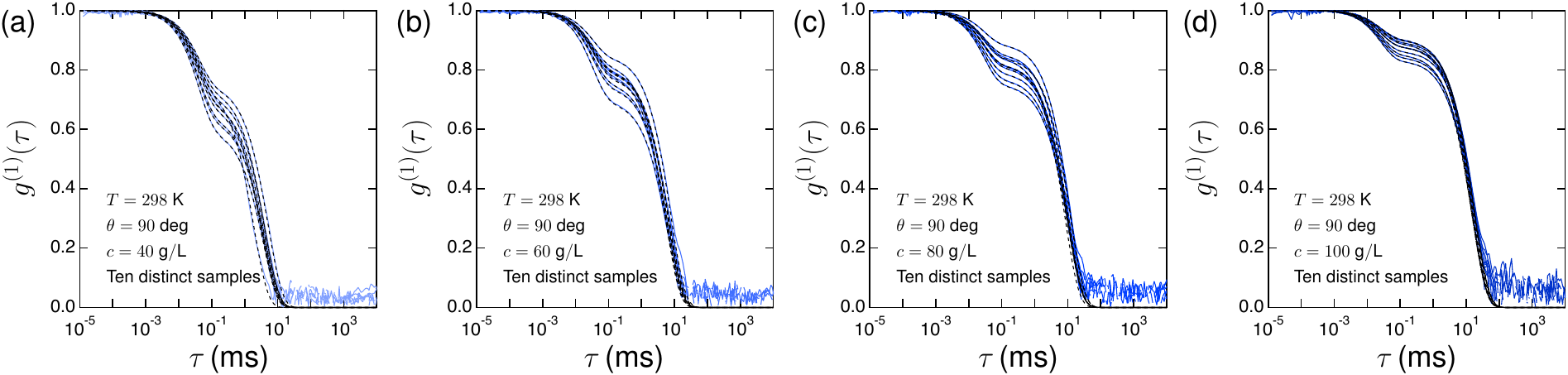}
\caption{The $g^{(1)}(\tau)$ profile measured at ten distinct samples prepared under identical conditions at $T=298$ K and $\theta=90^{\circ{}}$ for (a) $c = 40$ g$/$L, (b) $c = 60$ g$/$L, (c) $c = 80$ g$/$L, and (d) $c = 100$ g$/$L.
Each dashed curve represents the least-squares fit to $g^{(1)}(\tau)$ with Eq.~(1) in the main text.
}
\label{fig:10samples}
\end{figure}

We demonstrate the non-reproducibility of the slow mode in distinct samples prepared under identical conditions.
Figure~\ref{fig:10samples}(a)--(d) shows the $g^{(1)}(\tau)$ profile measured for ten distinct samples for $c=40$--$100$ g$/$L at $T=298$ K and $\theta=90^{\circ{}}$, indicating that the amplitude of the slow mode explicitly depends on the sample, which is known as the characteristic feature of the slow mode \cite{Ho2002_slowmode}.
Using $a_f$ obtained from the fit by Eq.~(1) in the main text (each black dashed curve), we resolve $R_{\theta,f}$ of each ten distinct sample shown in Fig.~2(b) in the main text (filled circles).
For the $\theta$ dependence at each $c$, one representative sample is chosen for the measurement, accounting for the sample-to-sample variation prepared under identical conditions.

\newpage

\section{All data on the $q^2$ dependence for $R_{\theta,f}$ and $R_{\theta,s}$}

\begin{figure}[h!]
\centering
\includegraphics[width=0.8\linewidth]{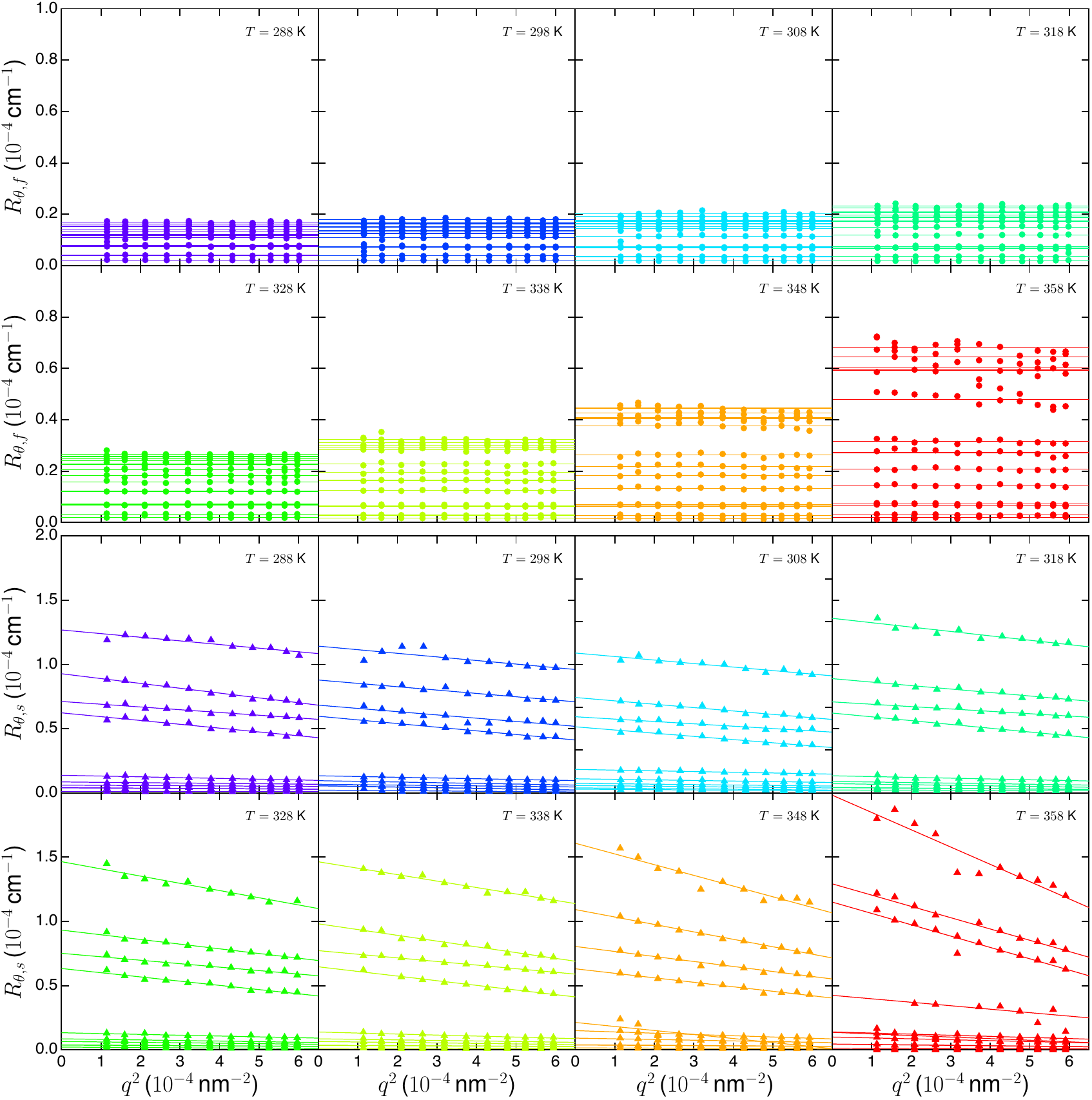}
\caption{The summary of Guinier plot for $R_{\theta,f}$ (upper panel) and $R_{\theta,s}$ (lower panel) over $T=288$--$358$ K.
Each line represents the least-squares fit to the data with a constant (upper panel) and with a linear function (lower panel).
}
\label{fig:R_f_s_q2}
\end{figure}

We summarize all data on the $q^2$ dependence of $R_{\theta,f}$ (upper panel) and $R_{\theta,s}$ (lower panel) over $T=288$--$358$ K in Fig.~\ref{fig:R_f_s_q2}.
The linear $q^2$ dependence indicates the validity of the approximation $qR_g \ll 1$, allowing a Guinier fit [$R_{\theta,s} \propto 1 - (R_g^2/3)q^2$], where $R_g$ is the radius of gyration.
At all $T$, $R_{\theta,f}$ is approximately independent of $q^2$ (black solid lines).
On the other hand, $R_{\theta,s}$ exhibits significant $q^2$ dependence (black solid lines), providing $R_g$ from the slope and intercept.
Because $R_{\theta,f}$ is approximately independent of $q^2$, the Guinier fit can also be applied to $R_{\theta}\,(=R_{\theta,f}+R_{\theta,s})$.

\newpage

\section{Comparison of virial expansion and Berry plot with universal EOS}

\begin{figure}[h!]
\centering
\includegraphics[width=0.3\linewidth]{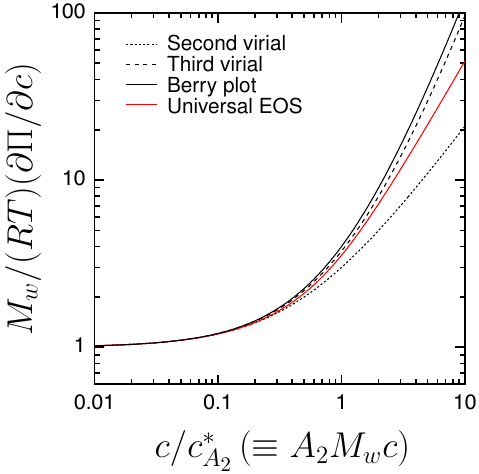}
\caption{Comparison of the universal EOS (red solid curve) with the virial expansion up to the second term (black fine dashed curve), up to the third term (black coarse dashed curve), and Berry plot (black solid curve).}
\label{fig:comp_berry_EOS}
\end{figure}

We assess how the virial expansion [Eq.~(4) in the main text] and the Berry plot deviate from the universal EOS (see End Matter in the main text) under good solvent conditions.
In good solvents, the second and third virial coefficients are related as $A_3/(A_{2}^2 M)=0.25$ \cite{Flory1953_polymer}.
Figure~\ref{fig:comp_berry_EOS} compares the universal EOS (red solid curve) with the virial expansion up to the second term (black fine dashed curve), up to the third term (black coarse dashed curve), and Berry plot (black solid curve).
In the dilute regime ($c < c^*_{A_2}$), the predictions of these equations converge to the universal EOS, but deviations from the universal EOS increase as the system enters the semidilute regime ($c > c^*_{A_2}$).
Therefore, the Berry plot is best suited for data in a sufficiently dilute regime.

\newpage

\section{Theta temperature $T_{\Theta}$ in aqueous PEG solution in prior studies}

\begin{figure}[h!]
\centering
\includegraphics[width=0.3\linewidth]{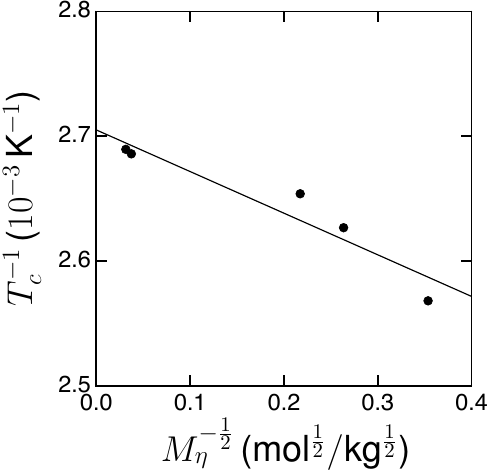}
\caption{Replot of $M_\eta^{-1/2}$ dependence of the inverse critical temperature $T_c^{-1}$ (LCST in aqueous PEG solutions), using $T_c$ reported previously \cite{Saeki1976_thetaTempLCST}.
Solid line represents the least-squares fit to the data with a linear function.}
\label{fig:CP_thetaTemp}
\end{figure}

To compare with the $\Theta$ temperature $T_{\Theta}$ evaluated by mode-resolved analysis in this study (see main text), we summarized the reference value of $T_{\Theta}$ of the aqueous PEG solution evaluated in prior studies.
The $T_{\Theta}$ has been evaluated through several methods: cloud point \cite{Saeki1976_thetaTempLCST,Boucher1976_thetaTemp_CP,Ataman1987_thetaTemp_CP,Fischer2000_thetaTempCPC}, SLS \cite{Strazielle1968_thetaTempSLS,William1983_SLSPEOTemp20to90,Venohr1998_thetaTempSLS}, X-ray scattering \cite{Pedersen2005_thetaTempSAXS}, and viscometry \cite{Amu1982_thetaTemp_viscosity} (summarized in Table~\ref{tab:T_theta}).
Here, for the data in Ref.~\cite{Saeki1976_thetaTempLCST}, we replot the reported lower critical solution temperature $T_c$ to evaluate $T_\theta$ against the viscosity-average molecular weight $M_\eta$, based on the relation of $T_c^{-1} - T_\Theta^{-1}\propto M^{-1/2}$ \cite{Flory1953_polymer}.
The least-squares fit to the data with a linear function (Solid line) in Fig.~\ref{fig:CP_thetaTemp} yields $T_\Theta = 369.7(19)$ K from the intercept of the black solid line ($T_c^{-1}$ at $M_\eta \to \infty$).

The $T_\Theta$ determined from the cloud points varies in the range of $369$--$370$ K, showing negligible dependence on the sample or analytical procedure.
In this study, we adopt $T_\Theta=369.5(11)$ K evaluated from the Shultz-Flory plot corrected for polydispersity \cite{Fischer2000_thetaTempCPC} as the most reliable value for comparison in the main text.
On the other hand, $T_\Theta$ determined from static light scattering (SLS) and X-ray scattering ranges from $350$--$376$ K, with deviations of up to approximately $20$ K from the cloud-point values.
These significant deviations may be attributed to the presence of unrecognized aggregates or nanobubbles in the solution.
The $T_\theta$ determined by the viscometry also shows the deviation from the cloud-point value.

\begin{table}[h!]
\caption{Comparison of $T_{\Theta}$ reported previously and evaluated in this study for the aqueous PEG solution.
}
\label{tab:T_theta}
\begin{ruledtabular}
\begin{tabular}{lll}
Reference 
& Method 
& Value
\\
\hline
\cite{Saeki1976_thetaTempLCST} 
& Cloud point (evaluated in Fig.~\ref{fig:CP_thetaTemp})
& $T_{\Theta}=369.7(19)$ K
\\
\cite{Boucher1976_thetaTemp_CP} 
& Cloud point ($M_n\approx 21$ kg$/$mol)
& $T_{\Theta}=369(3)$ K
\\
\cite{Ataman1987_thetaTemp_CP} 
& Cloud point ($M_n\approx 20$ kg$/$mol)
& $T_{\Theta}=369(2)$ K
\\
\cite{Fischer2000_thetaTempCPC} 
& Cloud point (Shultz-Flory plot corrected for polydispersity)
& $T_{\Theta}=369.5(11)$ K
\\
\hline
\cite{William1983_SLSPEOTemp20to90} 
& SLS ($M_w\approx 17$ kg$/$mol)
& $T_{\Theta}\approx 375$ K
\\
\cite{Strazielle1968_thetaTempSLS} 
& SLS ($M_w\approx 11.5$ kg$/$mol)
& $T_{\Theta}=370.2(18)$ K
\\
\cite{Strazielle1968_thetaTempSLS} 
& SLS ($M_w\approx 32.8$ kg$/$mol)
& $T_{\Theta}=375.2(25)$ K
\\
\cite{Venohr1998_thetaTempSLS} 
& SLS ($M_w\approx 13$ kg$/$mol)
& $T_{\Theta}=367.1(31)$ K
\\
\cite{Venohr1998_thetaTempSLS} 
& SLS ($M_w\approx 16$ kg$/$mol)
& $T_{\Theta}=350.3(20)$ K
\\
This study
& SLS ($M_w\approx 45.9$ kg$/$mol, see Sec.~S15)
& $T_{\Theta}=357.6(36)$ K
\\
This study
& Mode-Resolved analysis ($M_w\approx 45.5$ kg$/$mol, see main text)
& $T_{\Theta}=371.9(9)$ K
\\
\hline
\cite{Pedersen2005_thetaTempSAXS} 
& X-ray scattering (virial analysis, $M_n\approx 4.6$ kg$/$mol)
& $T_{\Theta}=373.1(10)$ K
\\
\cite{Pedersen2005_thetaTempSAXS} 
& X-ray scattering (Flory--Huggins theory analysis, $M_n\approx 4.6$ kg$/$mol)
& $T_{\Theta}=358.3(18)$ K
\\
\hline
\cite{Amu1982_thetaTemp_viscosity} 
& Viscometry (Stockmayer–Fixman plot)
& $T_{\Theta}\approx381.3$ K
\\
\end{tabular} 
\end{ruledtabular}
\end{table}

\newpage

\section{Evaluation of $M_w$ and $A_2$ using
the unresolved $R_\theta$}

\begin{figure}[h!]
\centering
\includegraphics[width=1\linewidth]{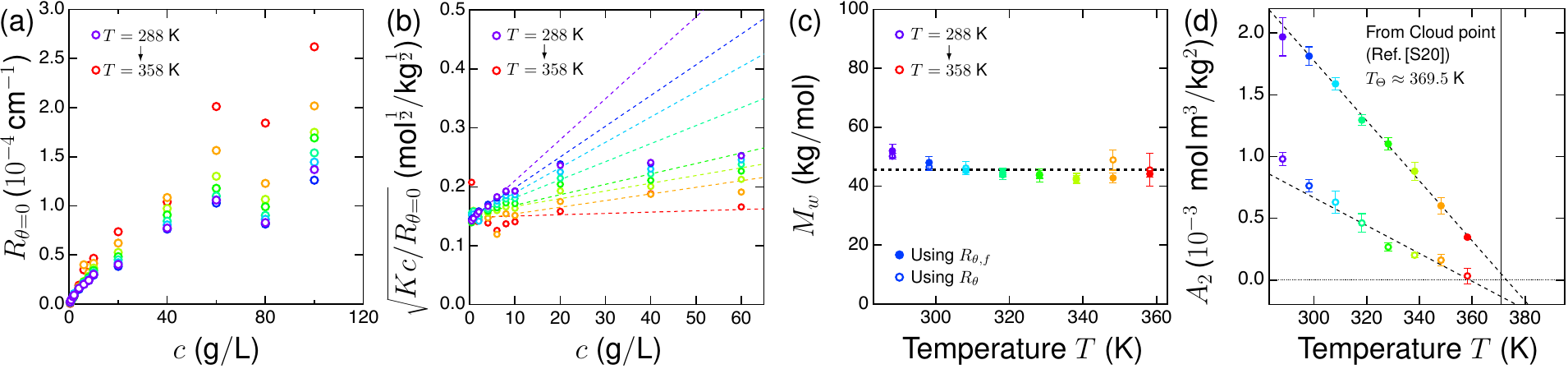}
\caption{Evaluation of $M_w$ and $A_2$ in the case of using the unresolved $R_{\theta}$.
(a) The $c$ dependence of $R_{\theta=0}$ over $T=288$--$358$ K.
(b) The Berry plot using $R_{\theta=0}$.
(c) Comparison of $T$ dependence of $M_w$ using $R_{\theta,f}$ (filled circles) and $R_{\theta}$ (open circles) obtained from the Berry plot.
From the least-squares fit to the data with a constant (black dashed line), we can obtain $M_w = 45.5(11)$ and $45.9(9)$ kg$/$mol for using $R_{\theta,f}$ and $R_{\theta}$, respectively.
(d) Comparison of $T$ dependence of $A_2$ using $R_{\theta,f}$ (filled circles) and $R_{\theta}$ (open circles) obtained from the Berry plot.
Each dashed line represents the least-squares fit to the data in the vicinity of $A_2 = 0$ with a linear function.
From the extrapolation of each dashed line to $A_2=0$, we obtain $T_{\Theta}= 371.9(9)$ K and $357.6(36)$ K for using $R_{\theta,f}$ and $R_{\theta}$, respectively.
The black solid line represents the $T_{\Theta}= 369.5(11)$ K determined from the cloud point in the previous report \cite{Fischer2000_thetaTempCPC}.
}
\label{fig:R_single}
\end{figure}

In the case of using the unresolved $R_{\theta=0}\,(=R_{\theta=0,f}+R_{\theta=0,s})$ over $T=288$--$358$ K [Fig.~\ref{fig:R_single}(a)], we evaluate $M_w$ and $A_2$ from the Berry plot [Fig.~\ref{fig:R_single}(b)].
Figure~\ref{fig:R_single}(c) and (d) summarize the comparison of $T$ dependence of $M_w$ and $A_2$ obtained by using $R_{\theta=0}$ (open circles) and $R_{\theta=0,f}$ (filled circles).
The $M_w$ evaluated from using $R_{\theta=0}$ and $R_{\theta=0,f}$ are both approximately independent of $T$, exhibiting $M_w= 45.9(9)$ kg$/$mol.
The $M_w$ obtained using $R_{\theta=0}$ is nearly identical to $M_w= 45.5(11)$ kg$/$mol obtained using $R_{\theta=0,f}$, because $a_f\approx 1$ in the dilute limit $(c \to 0)$.
In contrast, by extrapolating the $T$ dependence to $A_2=0$, $T_{\Theta}= 357.6(36)$ K is determined using $R_{\theta=0}$, which is explicitly smaller than that of $T_{\Theta}= 371.9(9)$ K using $R_{\theta=0,f}$ and $T_\Theta = 369.5(11)$ K determined from the cloud point in the previous report \cite{Fischer2000_thetaTempCPC}.
Here, we used the data in the vicinity of $A_2=0$ for the linear fit to extrapolate to $A_2 = 0$.

